\documentclass[12pt,english]{article}
\usepackage{times}
\usepackage[T1]{fontenc}
\usepackage{geometry}
\usepackage{graphicx}
\usepackage{setspace}
\usepackage{amssymb}

\makeatletter

\providecommand{\tabularnewline}{\\}

\usepackage{bm}

\usepackage{babel}
\makeatother
\begin{document}

\section*{\noindent Unconventional Array Design in the Autocorrelation Domain
- Isophoric \emph{1D} Thinning}

\noindent ~

\noindent \vfill

\noindent L. Poli,$^{(1)(2)}$ \emph{Senior Member, IEEE}, G. Oliveri,$^{(1)(2)}$
\emph{Fellow, IEEE}, N. Anselmi,$^{(1)(2)}$ \emph{Senior Member,
IEEE}, A. Benoni, $^{(1)}$ \emph{Member, IEEE}, L. Tosi, $^{(1)}$
\emph{Member, IEEE}, and A. Massa,$^{(1)(2)(3)(4)(5)}$ \emph{Fellow,
IEEE}

\noindent \vfill

\noindent {\footnotesize ~}{\footnotesize \par}

\noindent {\footnotesize $^{(1)}$} \emph{\footnotesize ELEDIA Research
Center} {\footnotesize (}\emph{\footnotesize ELEDIA}{\footnotesize @}\emph{\footnotesize UniTN}
{\footnotesize - University of Trento)}{\footnotesize \par}

\noindent {\footnotesize DICAM - Department of Civil, Environmental,
and Mechanical Engineering}{\footnotesize \par}

\noindent {\footnotesize Via Mesiano 77, 38123 Trento - Italy}{\footnotesize \par}

\noindent \textit{\emph{\footnotesize E-mail:}} {\footnotesize \{}\emph{\footnotesize lorenzo.poli,
giacomo.oliveri, nicola.anselmi.1, arianna.benoni, luca.tosi-1, andrea.massa}{\footnotesize \}@}\emph{\footnotesize unitn.it;}{\footnotesize \par}

\noindent {\footnotesize Website:} \emph{\footnotesize www.eledia.org/eledia-unitn}{\footnotesize \par}

\noindent {\footnotesize $^{(2)}$} \emph{\footnotesize CNIT - \char`\"{}University
of Trento\char`\"{} ELEDIA Research Unit }{\footnotesize \par}

\noindent {\footnotesize Via Mesiano 77, 38123 Trento - Italy}{\footnotesize \par}

\noindent {\footnotesize Website:} \emph{\footnotesize www.eledia.org/eledia-unitn}{\footnotesize \par}

\noindent {\footnotesize $^{(3)}$} \emph{\footnotesize ELEDIA Research
Center} {\footnotesize (}\emph{\footnotesize ELEDIA}{\footnotesize @}\emph{\footnotesize UESTC}
{\footnotesize - UESTC)}{\footnotesize \par}

\noindent {\footnotesize School of Electronic Science and Engineering,
Chengdu 611731 - China}{\footnotesize \par}

\noindent \textit{\emph{\footnotesize E-mail:}} \emph{\footnotesize andrea.massa@uestc.edu.cn}{\footnotesize \par}

\noindent {\footnotesize Website:} \emph{\footnotesize www.eledia.org/eledia}{\footnotesize -}\emph{\footnotesize uestc}{\footnotesize \par}

\noindent {\footnotesize $^{(4)}$} \emph{\footnotesize ELEDIA Research
Center} {\footnotesize (}\emph{\footnotesize ELEDIA@TSINGHUA} {\footnotesize -
Tsinghua University)}{\footnotesize \par}

\noindent {\footnotesize 30 Shuangqing Rd, 100084 Haidian, Beijing
- China}{\footnotesize \par}

\noindent {\footnotesize E-mail:} \emph{\footnotesize andrea.massa@tsinghua.edu.cn}{\footnotesize \par}

\noindent {\footnotesize Website:} \emph{\footnotesize www.eledia.org/eledia-tsinghua}{\footnotesize \par}

\noindent {\small $^{(5)}$} {\footnotesize School of Electrical Engineering}{\footnotesize \par}

\noindent {\footnotesize Tel Aviv University, Tel Aviv 69978 - Israel}{\footnotesize \par}

\noindent \textit{\emph{\footnotesize E-mail:}} \emph{\footnotesize andrea.massa@eng.tau.ac.il}{\footnotesize \par}

\noindent {\footnotesize Website:} \emph{\footnotesize https://engineering.tau.ac.il/}{\footnotesize \par}

\noindent \vfill

\noindent \emph{This work has been submitted to the IEEE for possible
publication. Copyright may be transferred without notice, after which
this version may no longer be accessible.}

\noindent \vfill

\newpage
\section*{Unconventional Array Design in the Autocorrelation Domain - Isophoric
\emph{1D} Thinning}

~

~

~

\begin{flushleft}L. Poli, G. Oliveri, N. Anselmi, A. Benoni, L. Tosi,
and A. Massa\end{flushleft}

\vfill

\begin{abstract}
\noindent The synthesis of thinned isophoric arrays (\emph{TIA}s)
radiating mask-constrained patterns is addressed. By leveraging on
the recently-introduced formulation of the design of antenna arrays
in the autocorrelation-domain (\emph{AD}), the \emph{TIA} synthesis
is recast as the matching of a target autocorrelation function derived
from the user-defined guidelines and objectives. By exploiting the
autocorrelation invariance of \emph{}cyclic binary sequences, the
\emph{AD} solution space is significantly reduced and it is efficiently
sampled by means of a discrete hybrid optimization approach. Two possible
implementations of the \emph{AD}-based \emph{TIA} formulation are
discussed and assessed in a set of representative numerical examples
concerned with both ideal and real radiators, which are full-wave
modeled to account for the mutual coupling effects. Comparisons with
traditional pattern-domain (\emph{PD}) synthesis methods are also
considered to point out the features and the advantages of \emph{AD}-based
approaches.

\vfill
\end{abstract}
\noindent \textbf{Key words}: Isophoric arrays, linear arrays, thinned
arrays, optimization methods.

\newpage
\section{Introduction and Motivations\label{sec:Introduction}}

\noindent Unconventional (antenna) arrays (\emph{UA}s) are emerging
as a key enabling technology across several applicative domains including
satellite and terrestrial communications, automotive radar and sensing,
far-field and near-field imaging, direction-of-arrival estimation,
and localization \cite{Rocca 2016}-\cite{Rocca 2022}. The success
of \emph{UA}s is motivated by the possibility to overcome the limitations
of traditional (i.e., conventional) architectures in terms of radiation
performance, flexibility, and fabrication complexity \cite{Rocca 2016}-\cite{Rocca 2022}.
Within \emph{UA}s, thinned isophoric arrays (\emph{TIA}s) \cite{Haupt 2010}\cite{Skolnik 1964.b}-\cite{Keizer 2008}
have been proposed as a competitive solution for the simultaneous
minimization of both the cost and the weight of the antenna system,
while still radiating suitable power patterns \cite{Rocca 2016}.
Moreover, \emph{TIA}s may potentially accommodate multiple co-existing
functionalities or services on the same physical antenna aperture
through interleaving \cite{Oliveri 2010.b}\cite{Oliveri 2014}.

\noindent Unfortunately, these advantages are yielded with more complex
mechanisms to control/adjust the features of the radiation pattern
with respect to other classes of both traditional/conventional and
unconventional array layouts \cite{Rocca 2016} because of (\emph{i})
the reduced number of degrees-of-freedom (\emph{DoF}s) and (\emph{ii})
the discrete, non-linear, and multi-minima nature of the associated
design/pattern control problem \cite{Rocca 2016}\cite{Oliveri 2011e}-\cite{Wolpert 1997}.

\noindent In the state-of-the-art literature, a wide variety of design
methods has been proposed to effectively control the pattern features
of \emph{TIA}s. Low-complexity strategies based on statistical techniques
have been initially investigated owing to their efficiency, also dealing
with very wide apertures, but they only allow the prediction/control
of a limited set of radiation features {[}e.g., the average sidelobe
level (\emph{SLL}){]} \cite{Skolnik 1964.b}\cite{Mailloux 1991}.
Evolutionary global optimization techniques have been exploited \cite{Haupt 2010}\cite{Haupt 1994}\cite{Bray 2002}\cite{Quevedo-Teruel 2006}-\cite{Ha 2016},
as well, but they may suffer from slow convergence and generally imply
a heavy computational burden when synthesizing large \emph{TIA}s \cite{Rocca 2009}.
The use of analytical techniques has been considered to circumvent
these issues, which are related to high-dimensional solution spaces
\cite{Oliveri 2009.a}\cite{Oliveri 2010}, as well as their hybridization
with optimization techniques to combine the advantages of both concepts
\cite{Oliveri 2011e}\cite{Salucci 2017}, but the beam control still
turned out to be limited (i.e., only the reduction of the \emph{SLL}
has been usually yielded) \cite{Oliveri 2009.a}\cite{Oliveri 2010}.

\noindent Recently, an innovative paradigm has been introduced that
reformulates the array synthesis problem from the pattern domain (\emph{PD})
to a transformed domain \cite{Oliveri 2020}, namely the \emph{autocorrelation
domain} (\emph{AD}) \cite{Oliveri 2020}, by taking advantage of the
analytical relationships between the autocorrelation of the array
excitations and the associated power pattern samples. Starting from
binary sequences with known redundancy properties, closed-form synthesis
expressions have been derived that \emph{a-priori} guarantee to obtain
\emph{TIA}s layouts fulfilling user-defined specifications on the
\emph{SLL}, the directivity, the half-power beam-width, and the power
pattern along user-defined directions \cite{Oliveri 2020}. Notwithstanding
the solid theoretical foundation, its efficiency, and its effectiveness,
such a \emph{TIA}-oriented design process \cite{Oliveri 2020} has
some basic limitations since it (\emph{i}) relies on the availability
of analytical sequences with known autocorrelation features, thus
it is severely constrained to a reduced set of geometries and thinning
factors \cite{Oliveri 2020}, and (\emph{ii}) does not allow the user
to define arbitrary beam shaping / sidelobe profiles, the pattern
properties being mostly dictated by the class of analytical sequences
at hand \cite{Oliveri 2020}.

\noindent By leveraging on the theoretical framework developed in
\cite{Oliveri 2020}, the objective of this work is to define an innovative
general-purpose synthesis method where the design of a \emph{TIA}s
affording user-defined beam-shapes is recast as the matching of suitable
autocorrelation functions without the need of auxiliary analytical
binary sequences. More specifically, the setup of \emph{TIA} layouts
matching the user-defined synthesis objectives (e.g., compliance with
a mask) is firstly reformulated as a combinatorial problem in the
\emph{AD}. By exploiting the cyclic periodicity property of the autocorrelation
function of discrete sequences \cite{Oliveri 2009.a}, the solution
space of the arising \emph{AD} problem is significantly reduced and
it is successively sampled with a hybrid global/local evolutionary
optimization technique.

\noindent To the best of the authors' knowledge, the novelty of our
work over existing literature includes (\emph{i}) the introduction
of a generalized formulation of \emph{TIA} synthesis problems in the
\emph{AD} that, unlike previous \emph{AD}-based approaches, allows
the designer to set arbitrary pattern constraints without the need
of auxiliary analytical binary sequences, by also reducing the dimensionality
of the solution space with respect to standard \emph{PD} techniques;
(\emph{ii}) the derivation of two innovative implementations of the
\emph{AD} formulation based on either the {}``Mask Equality'' (\emph{ME})
strategy or the {}``Feasible Pattern Equality\emph{''} (\emph{FPE})
one.

\noindent The outline of the paper is as follows. The formulation
of the \emph{TIA} synthesis problem in the \emph{AD} is discussed
in Sect. \ref{sec:Problem-Formulation}, while Section \ref{sec:Method}
details the solution process based on an evolutionary optimizer. Afterwards,
a representative set of numerical experiments is illustrated to assess
the features and the potentialities of the proposed \emph{TIA} synthesis
method also in comparison with state-of-the-art thinning techniques
(Sect. \ref{sec:Numerical-Analysis-and}). Conclusions and final remarks
follow (Sect. \ref{sec:Conclusions-and-Remarks}).

\section{\noindent \emph{TIA} Synthesis Formulation\label{sec:Problem-Formulation}}

\noindent Let us consider a linear \emph{TIA} of $N$ elements {[}Fig.
1(\emph{a}){]} arranged along the $\widehat{\mathbf{z}}$ direction
in a regular grid of $P$ ($P>N$) candidate locations, \{$\left(x_{p},y_{p},z_{p}\right)=\left(0,0,d_{p}\right)$;
$p=0,...,P-1$\}, where\begin{equation}
d_{p}=\left(p-\frac{P-1}{2}\right)\times\Delta z,\label{eq: candidate locations}\end{equation}
$\Delta z$ being the grid step. Such a \emph{TIA} is univocally described
by the \emph{TIA descriptor vector} $\bm{\alpha}$ {[}$\bm{\alpha}\in\mathcal{D}\left(\bm{\alpha}\right)$,
$\mathcal{D}\left(\bm{\alpha}\right)$ being the solution space{]}
whose $p$-th ($p=0,...,P-1$) entry, $\alpha_{p}$, is a binary value
($\alpha_{p}=1/0$ if the $p$-th grid location is occupied/empty)
and $N=\sum_{p=0}^{P-1}\alpha_{p}$. Moreover, the radiated power
pattern \cite{Mailloux 2005}%
\footnote{\noindent For the sake of notation simplicity, single-polarization
antennas are hereinafter considered. However, double polarization
radiators can be straightforwardly taken into account within the same
theoretical framework by introducing co/cross polar masks.%
} turns out to be\begin{equation}
\mathcal{E}\left(u;\bm{\alpha}\right)\triangleq\left|\sum_{p=0}^{P-1}\alpha_{p}F_{p}\left(u\right)\exp\left(j\frac{2\pi d_{p}}{\lambda}u\right)\right|^{2}\label{eq:pattern}\end{equation}
where $\lambda$ is the wavelength and $F_{p}\left(u\right)$ is the
$p$-th ($p=0,...,P-1$) embedded element pattern.

\noindent According to this notation, the problem of synthesizing
a \emph{TIA} with mask-constrained beam features can be then stated
in the (traditional) \emph{PD} as follows

\begin{quotation}
\noindent \emph{Pattern-Domain TIA} (\emph{PD-TIA}) \emph{Design Problem}
- Find $\bm{\alpha}^{opt}$ {[}$\bm{\alpha}^{opt}\in\mathcal{D}\left(\bm{\alpha}\right)${]}
such that\begin{equation}
\widetilde{\mathcal{E}}\left(u;\bm{\alpha}^{opt}\right)\leq\mathcal{M}\left(u\right)\label{eq:mask constraint}\end{equation}
within the visible range $u\in\left[-1,1\right]$.
\end{quotation}
\noindent In (\ref{eq:mask constraint}), $\widetilde{\mathcal{E}}\left(u;\bm{\alpha}\right)$
is the \emph{TIA} normalized power pattern,

\noindent \begin{equation}
\widetilde{\mathcal{E}}\left(u;\bm{\alpha}\right)\triangleq\frac{\mathcal{E}\left(u;\bm{\alpha}\right)}{\mathcal{E}\left(0;\bm{\alpha}\right)},\label{eq:normalized pattern}\end{equation}
$u=\cos\theta$, and $\mathcal{M}\left(u\right)$ is the user-defined
target pattern mask.

\noindent The \emph{PD-TIA} formulation has been traditionally adopted
in the state-of-the-art literature by setting uniform-sidelobe target
masks \cite{Haupt 1994}\cite{Oliveri 2011e} and the resulting optimization
problem, often formulated as the minimization of the \emph{mask matching
error},\begin{equation}
\xi\left(\bm{\alpha}\right)\triangleq\frac{\int_{u=-1}^{1}\Xi\left[\widetilde{\mathcal{E}}\left(u;\bm{\alpha}\right)-\mathcal{M}\left(u\right)\right]\mathrm{d}u}{\int_{u=-1}^{1}\mathcal{M}\left(u\right)\mathrm{d}u},\label{eq:mask matching}\end{equation}
 $\Xi\left[\cdot\right]$ being the step \emph{}function, has been
addressed with several and effective strategies \cite{Haupt 1994}\cite{Keizer 2008}\cite{Oliveri 2011e}
\cite{Quevedo-Teruel 2006}\cite{Deligkaris 2009}\cite{Gu 2019}\cite{Wang 2014}.
However, the performance of the methods formulated within the \emph{PD}
framework are generally limited by the computation of (\ref{eq:pattern})
for each trial solution. This may result in a non-negligible computational
burden especially for large arrangements owing to the need to sample
the power pattern at a sufficiently-fine angular resolution. Moreover,
the \emph{PD-TIA} formulation does not take advantage of the cyclic
shift property of binary sequences \cite{Oliveri 2009.a}\cite{Leeper 1999},
which is effectively exploited by analytical design techniques \cite{Oliveri 2009.a}\cite{Oliveri 2011e}\cite{Oliveri 2020}\cite{Leeper 1999}
to reduce the search space as well as to improve its sampling.

\noindent Starting from these considerations and leveraging on the
theoretical framework introduced in \cite{Oliveri 2020}, an alternative
formulation of the \emph{TIA} design problem is proposed hereinafter
by reformulating the user-defined objectives from the \emph{PD} to
the \emph{AD}. Towards this end, the power pattern expression in (\ref{eq:pattern})
is firstly rewritten \cite{Oliveri 2020} as\begin{equation}
\mathcal{E}\left(u;\bm{\alpha}\right)=\left|F\left(u\right)\right|^{2}\left|\sum_{k=0}^{P-1}\sqrt{\Gamma_{k}\left(\bm{\alpha}\right)}\exp\left(j\psi_{k}\left(\bm{\alpha}\right)\right)\times\mathcal{S}\left(\frac{2\pi\Delta zu}{\lambda}-\frac{2\pi k}{P}\right)\right|^{2}\label{eq:pattern in autocorrelation}\end{equation}
by using the \emph{large array approximation} \cite{Mailloux 2005}
(i.e., $F_{p}\left(u\right)=F\left(u\right)$, $p=0,...,P-1$).

\noindent In (\ref{eq:pattern in autocorrelation}), $\Gamma_{k}\left(\bm{\alpha}\right)$
is the $k$-th ($k=0,...,P-1$) element\begin{equation}
\Gamma_{k}\left(\bm{\alpha}\right)\triangleq\sum_{s=0}^{P-1}\left[\gamma_{s}\left(\bm{\alpha}\right)\exp\left(j\frac{2\pi s}{P}k\right)\right]\label{eq:dft autocorr}\end{equation}
of the discrete Fourier transform (\emph{DFT}) of the \emph{autocorrelation}
of the \emph{TIA} descriptors, $\bm{\gamma}\left(\bm{\alpha}\right)$
$\triangleq$ \{$\gamma_{s}\left(\bm{\alpha}\right)$; $s=0,...,P-1$\},
whose $s$-th entry is given by

\noindent \begin{equation}
\gamma_{s}\left(\bm{\alpha}\right)\triangleq\sum_{p=0}^{P-1}\alpha_{p}\alpha_{\left\lfloor p+s\right\rfloor _{P}},\label{eq:autocorrelation}\end{equation}
$\left\lfloor \cdot\right\rfloor _{P}$ being the modulo-$P$ operator.
Moreover, $\psi_{k}\left(\bm{\alpha}\right)$ is a deterministic term
equal to the phase of the $k$-th coefficient of the \emph{DFT} of
the descriptor sequence (i.e., $\psi_{k}\left(\bm{\alpha}\right)\triangleq\angle\left\{ \sum_{p=0}^{P-1}\left[\alpha_{p}\exp\left(j\frac{2\pi p}{P}k\right)\right]\right\} $),
while $\mathcal{S}\left(\nu\right)$ is the 1D interpolation function
defined as\begin{equation}
\mathcal{S}\left(\nu\right)\triangleq\frac{\sin\left(\frac{P}{2}\times\nu\right)}{P\sin\left(\frac{\nu}{2}\right)}\exp\left(j\frac{P-1}{2}\times\nu\right).\label{eq:interpolation function}\end{equation}
By sampling (\ref{eq:pattern in autocorrelation}) at $u=u_{k}=\frac{\lambda}{P\Delta z}k$
($k=0,...,P-1$) and using the properties of the interpolation function
\cite{Oliveri 2020}, the mask-matching condition in (\ref{eq:mask constraint})
can be rewritten as follows\begin{equation}
\Gamma_{k}\left(\bm{\alpha}\right)\leq N^{2}\times\mathcal{M}\left(\frac{\lambda}{P\Delta z}k\right)\label{eq:sampled constraint}\end{equation}
($k=0,...,P-1$) since $\widetilde{\mathcal{E}}\left(u_{k};\bm{\alpha}\right)=\frac{\Gamma_{k}\left(\bm{\alpha}\right)}{N^{2}}$
and $\mathcal{E}\left(0;\bm{\alpha}\right)=\left|F\left(u\right)\right|^{2}\times N^{2}$
\cite{Oliveri 2020} ($\mathcal{E}\left(0;\bm{\alpha}\right)\triangleq\left|F\left(u\right)\right|^{2}\times\left|\sum_{p=0}^{P-1}\alpha_{p}\right|^{2}$).

\noindent It is worth remarking that, even though still formulated
in the \emph{PD}, (\ref{eq:sampled constraint}) highlights the relationship
between the \emph{DFT} of the autocorrelation of $\bm{\alpha}$ (on
the left-hand) and the user-defined mask requirements (on the right-hand).
To formulate the \emph{TIA} synthesis in the \emph{AD}, it would be
in principle easy to apply the inverse \emph{DFT} (\emph{IDFT}) operator
directly to (\ref{eq:sampled constraint}), but this is not possible
since the Fourier Transform is not a monotonic operator and it cannot
be applied to inequalities such as that in (\ref{eq:sampled constraint}).
To circumvent such an issue, the solution domain is bounded to the
descriptors $\bm{\alpha}$ that fulfil the \emph{equality-restricted}
version of (\ref{eq:sampled constraint}):\begin{equation}
\Gamma_{k}\left(\bm{\alpha}\right)=N^{2}\times\mathcal{M}\left(\frac{\lambda}{P\Delta z}k\right)\label{eq:restricted equality}\end{equation}
($k=0,...,P-1$). Thanks to this assumption, the \emph{TIA} design
problem can be formulated in the \emph{AD} as follows

\begin{quotation}
\noindent \emph{Mask-Equality Autocorrelation-Domain TIA} (\emph{ME-AD-TIA})
\emph{Design Problem} - Find $\bm{\alpha}^{opt}$ {[}$\bm{\alpha}^{opt}\in\mathcal{D}\left(\bm{\alpha}\right)${]}
such that the condition

\noindent \begin{equation}
\gamma_{s}\left(\bm{\alpha}^{opt}\right)=N^{2}\times\mu_{s}\left(\mathbf{M}\right)\label{eq:matching autocorrelation mask}\end{equation}
($s=0,...,P-1$) is fulfilled, $\mathbf{M}\triangleq\left\{ M_{k}\triangleq\mathcal{M}\left(\frac{\lambda}{P\Delta z}k\right);\, k=0,...,P-1\right\} $
being the set of $P$ mask samples, while\begin{equation}
\mu_{s}\left(\mathbf{M}\right)\triangleq\frac{1}{P}\sum_{k=0}^{P-1}\left[M_{k}\exp\left(-j\frac{2\pi k}{P}s\right)\right]\label{eq:idft mask}\end{equation}
is the corresponding $s$-th ($s=0,...,P-1$) \emph{IDFT} coefficient.
\end{quotation}
\noindent Unfortunately this \emph{AD} formulation has some drawbacks
since strictly enforcing (\ref{eq:matching autocorrelation mask})
is not actually equivalent to fulfill (\ref{eq:sampled constraint}).
First of all, such a reformulation may result in a physically unfeasible
problem. As a matter of fact, there is no guarantee that there is
a \emph{TIA} that radiates a pattern matching an - even though realistic
- arbitrary user-defined mask $\mathcal{M}\left(u\right)$ in a discrete,
but uniform, set of angular points. Otherwise, sub-optimal \emph{TIA}
layouts may be synthesized owing to the over-constrained nature of
the design problem at hand. 

\noindent Therefore, an alternative \emph{AD}-based approach is introduced.
More specifically, instead of directly fulfilling the mask constraints,
the \emph{TIA} descriptors are set to match the autocorrelation function
of an \emph{auxiliary} fully-populated array (\emph{AFPA}) with excitations
$\mathbf{w}^{feas}\triangleq\left\{ w_{p}^{feas};\, p=0,...,P-1\right\} $
whose pattern complies with the user-defined (realistic) mask $\mathcal{M}\left(u\right)$:\begin{equation}
\mathbf{w}^{feas}=\arg\left\{ \left.\max_{\mathbf{w}}\left[\mathcal{E}\left(0;\mathbf{w}\right)\right]\right|\mathcal{E}\left(u;\mathbf{w}\right)\leq\mathcal{M}\left(u\right)\right\} .\label{eq:CP Problem}\end{equation}
This problem is now always feasible, the target autocorrelation being
that of a fully-populated array fulfilling (\ref{eq:mask constraint}).
Moreover, since (\ref{eq:CP Problem}) is a constrained convex minimization
problem, it can be solved by conventional deterministic techniques
\cite{Rocca 2016}\cite{Sartori 2013}\cite{DUrso 2007}.

\noindent Mathematically, the arising \emph{TIA} synthesis problem
can be then stated as

\begin{quotation}
\noindent \emph{Feasible-Pattern-Equality Autocorrelation-Domain TIA}
(\emph{FPE-AD-TIA}) \emph{Design Problem} - Find $\bm{\alpha}^{opt}$
{[}$\bm{\alpha}^{opt}\in\mathcal{D}\left(\bm{\alpha}\right)${]} such
that

\noindent \begin{equation}
\gamma_{s}\left(\bm{\alpha}^{opt}\right)=N^{2}\times\mu_{s}\left(\mathbf{E}^{feas}\right)\label{eq:matching feasible pattern}\end{equation}
($s=0,...,P-1$) holds true.
\end{quotation}
\noindent In (\ref{eq:matching feasible pattern}), $\mathbf{E}^{feas}$
consists of the samples of the auxiliary pattern $\mathcal{E}^{feas}\left(u\right)$
{[}$\mathcal{E}^{feas}\left(u\right)\triangleq\mathcal{E}\left(u;\mathbf{w}^{feas}\right)${]}\begin{equation}
\mathbf{E}^{feas}\triangleq\left\{ E_{k}^{feas}\triangleq\mathcal{E}^{feas}\left(\frac{\lambda}{P\Delta z}k\right);\, k=0,...,P-1\right\} \label{eq:}\end{equation}
and the $s$-th ($s=0,...,P-1$) \emph{feasible} \emph{autocorrelation}
coefficient, $\mu_{s}\left(\mathbf{E}^{feas}\right)$, is equal to\begin{equation}
\mu_{s}\left(\mathbf{E}^{feas}\right)\triangleq\frac{1}{P}\sum_{k=0}^{P-1}\left[\mathcal{E}^{feas}\left(\frac{\lambda}{P\Delta z}k\right)\exp\left(-j\frac{2\pi k}{P}s\right)\right].\label{eq:W Rsciugni}\end{equation}
Unlike the \emph{ME-AD-TIA} formulation, the \emph{FPE-AD-TIA} one
yields to a two-step solution process where (\emph{i}) an \emph{AFPA}
is firstly derived by solving (\ref{eq:CP Problem}) through a state-of-the-art
convex programming technique \cite{Rocca 2016}\cite{Sartori 2013}
and (\emph{ii}) the actual \emph{TIA} design step is carried out afterwards
in the \emph{AD} according to (\ref{eq:matching feasible pattern}).
Since several efficient algorithms are available to address the auxiliary
step (\emph{i}), \emph{}including the popular Matlab subroutine \emph{fmincon}
\cite{Rocca 2016}\cite{Sartori 2013}\cite{DUrso 2007}, only the
step (\emph{ii}) with the solution of (\ref{eq:matching feasible pattern})
will be detailed in the following. 

\noindent It is worthwhile to point out that the conceived autocorrelation-driven
paradigm is not limited to the constraints stated in (\ref{eq:matching autocorrelation mask})
or (\ref{eq:matching feasible pattern}), but alternative objectives
may be seamlessly taken into account within the same framework by
exploiting (\ref{eq:W Rsciugni}), thus defining an \emph{AFPA} that
fulfils (even) non pattern-related constraints (e.g., capacity maximization).

\section{\noindent Evolutionary-Based \emph{TIA} Synthesis\label{sec:Method}}

\noindent The solution of the \emph{TIA} synthesis problem, $\bm{\alpha}^{opt}$
{[}$\bm{\alpha}^{opt}\in\mathcal{D}\left(\bm{\alpha}\right)${]},
in the \emph{AD} for both the \emph{ME-AD-TIA} (\ref{eq:matching autocorrelation mask})
and the \emph{FPE-AD-TIA} (\ref{eq:matching feasible pattern}) formulations
(analogously to that in the \emph{PD}) requires a suitable search
procedure to efficiently/effectively sample the solution space $\mathcal{D}\left(\bm{\alpha}\right)$.
Indeed, (\emph{a}) the relation between the \emph{DoF}s of the \emph{TIA}
design, $\bm{\alpha}$, and the corresponding synthesis objective
is highly non-linear by definition of the autocorrelation function
(\ref{eq:autocorrelation}), and (\emph{b}) both the dimensionality
and the cardinality of $\mathcal{D}\left(\bm{\alpha}\right)$ of a
realistic \emph{TIA} design problem may be very huge, $P$ being in
the order of hundreds. Moreover, (\emph{c}) the binary nature of the
unknowns suggests to consider a binary-variable global optimizer as
solver tool \cite{Rocca 2009}.

\noindent However, just addressing the problem solution as a standard
binary optimization may be sub-optimal since it would neglect some
analytic properties of the autocorrelation function (\ref{eq:autocorrelation}).
For instance, it is known from combinatorial theory \cite{Oliveri 2009.a}\cite{Oliveri 2020}
that a $\sigma$-positions cyclically-shifted version of any \emph{parent}
sequence $\bm{\alpha}$, $\bm{\alpha}^{\left(\sigma\right)}$, which
is defined as\begin{equation}
\bm{\alpha}^{\left(\sigma\right)}\triangleq\left\{ \alpha_{\left\lfloor p+\sigma\right\rfloor _{P}};\, p=0,...,P-1\right\} ,\label{eq:shifted}\end{equation}
exhibits the same autocorrelation function of the generating sequence
(\emph{invariance property}), that is\begin{equation}
\gamma_{s}\left(\bm{\alpha}^{\left(\sigma\right)}\right)\equiv\gamma_{s}\left(\bm{\alpha}\right)\label{eq:autocorr shift invariance}\end{equation}
($s=0,...,P-1$). The exploitation of such a property (\ref{eq:autocorr shift invariance})
in the \emph{AD} problem formulations can play a significant role
in efficiently sampling $\mathcal{D}\left(\bm{\alpha}\right)$ towards
$\bm{\alpha}^{opt}$. As a matter of fact and unlike conventional
\emph{PD}-based strategies, the size of $\mathcal{D}\left(\bm{\alpha}\right)$
can be considerably {}``shrinked'' by taking into account that the
set of cyclically-shifted versions of a parent \emph{}sequence (i.e.,
\{$\bm{\alpha}^{\left(\sigma\right)}$; $\sigma=0,...,P-1$\}) has
the same autocorrelation function.

\noindent Accordingly, the global binary optimizer adopted for solving
the \emph{AD}-based \emph{TIA} problems is only responsible of generating
the \emph{parent} sequences, while the actual optimal $\bm{\alpha}^{opt}$
is afterwards found by simply comparing the members of the selected
$\bm{\alpha}$-family {[}i.e., directly checking the $P$ shifted
sequences generated with (\ref{eq:shifted}){]}. In particular, a
genetic algorithm (\emph{GA})-based mechanism \cite{Haupt 1994}\cite{Oliveri 2011e}\cite{Salucci 2017}\cite{Rocca 2009}
is used to evolve the parents sequences owing to (\emph{a}) its well-known
effectiveness in solving discrete problems with highly non-linear
and multi-minima cost functions to be minimized/maximized \cite{Rocca 2009},
(\emph{b}) the excellent results in dealing with standard \emph{TIA}
syntheses \cite{Haupt 1994}\cite{Oliveri 2011e}\cite{Salucci 2017},
and (\emph{c}) the possibility to easily include \emph{a-priori} knowledge
and/or design constraints by suitably adjusting the evolutionary operators
\cite{Oliveri 2011e}\cite{Salucci 2017}. More in detail, the proposed
\emph{GA}-based technique implements the following customizations
with respect to the state-of-the-art \emph{GA}s \cite{Haupt 1994}\cite{Oliveri 2011e}\cite{Salucci 2017}:

\begin{itemize}
\item \noindent \textbf{Initialization} ($i=0$) - As for the \emph{ME-AD-TIA}
formulation, use a standard random initialization by setting the $q$-th
($q=0,...,Q-1$; $Q$ being the population size) trial solution $\bm{\alpha}^{\left(q\right)}$
to\begin{equation}
\bm{\alpha}_{0}^{\left(q\right)}\triangleq\left\{ \alpha_{p0}^{\left(q\right)}=r_{p}^{\left(q\right)};\, p=0,...,P-1\right\} ,\label{eq:initialization ME}\end{equation}
$r_{p}^{\left(q\right)}$ being a random binary digit, while exploit
the auxiliary excitation set $\mathbf{w}^{feas}$ for the \emph{FPE-AD-TIA}
implementation as follows\begin{equation}
\bm{\alpha}_{0}^{\left(q\right)}\triangleq\left\{ \alpha_{p0}^{\left(q\right)}=\mathcal{R}\left[w_{\left\lfloor p+q\right\rfloor _{P}}^{feas}\right];\, p=0,...,P-1\right\} \label{eq:initialization FP}\end{equation}
($q=0,...,Q-1$) $\mathcal{R}\left[\cdot\right]$ being the rounding
operator (i.e., $\mathcal{R}\left[\cdot\right]=1$ if $\left[\cdot\right]\geq0.5$,
$\mathcal{R}\left[\cdot\right]=0$ otherwise); 
\item \textbf{Cost Function} - Define the cost function, which quantifies
the $\bm{\alpha}$-solution optimality, as\begin{equation}
\Phi\left(\bm{\alpha}\right)\triangleq\frac{\sum_{s=0}^{P-1}\left|\gamma_{s}\left(\bm{\alpha}\right)-\gamma_{s}^{*}\right|^{2}}{P}\label{eq:AD cost function}\end{equation}
by setting $\gamma_{s}^{*}$ ($s=0,...,P-1$) to $\gamma_{s}^{*}=N^{2}\times\mu_{s}\left(\mathbf{M}\right)$
and $\gamma_{s}^{*}=N^{2}\times\mu_{s}\left(\mathbf{E}^{feas}\right)$
for the \emph{ME-AD-TIA} formulation and the \emph{FPE-AD-TIA} one,
respectively;
\item \textbf{\emph{GA}} \textbf{Termination} - Stop the iterative process
of generating the $i$-th ($i=1,...,I$; $I$ being the maximum number
of iterations) population of trial solutions, \{$\bm{\alpha}_{i}^{\left(q\right)}$;
$q=0,...,Q-1$\}, when either $i=I$ ($\to$ $i_{conv}=I$) or the
following \emph{stagnation condition}\begin{equation}
\left|\Phi_{l}^{best}-\frac{1}{L}\sum_{i=l-L}^{l-1}\Phi_{i}^{best}\right|\leq\Phi_{th}\label{eq:Ciampufreggiu}\end{equation}
holds true ($\to$ $i_{conv}=i$). In (\ref{eq:Ciampufreggiu}), $\Phi_{th}$
is the user-defined convergence threshold, $\Phi_{i}^{best}$ is the
cost function value of the best guess found until the $i$-th ($i=1,...,I$)
iteration, and $L$ is the length of the stagnation window. Deduce
the optimal \emph{parent} sequence $\widehat{\bm{\alpha}}$ as $\widehat{\bm{\alpha}}$
$=$ $\arg$ \{ $\min_{q=0,...,Q-1}$ $\min_{i=0,...,I-1}$ $\left[\Phi\left(\bm{\alpha}_{i}^{\left(q\right)}\right)\right]$
\};
\item \textbf{Post}\textbf{\emph{-GA}} \textbf{Cyclic Shift} - Output the
final \emph{TIA} layout, $\bm{\alpha}^{opt}$, by applying the cyclic
$\sigma$-shift procedure to the optimal parent sequence $\widehat{\bm{\alpha}}$
and select that one with the minimum value of the mask matching error
$\xi\left(\widehat{\bm{\alpha}}^{\left(\sigma\right)}\right)$

\begin{equation}
\bm{\alpha}^{opt}\triangleq\arg\left\{ \min_{\sigma=0,...,P-1}\left[\xi\left(\widehat{\bm{\alpha}}^{\left(\sigma\right)}\right)\right]\right\} .\label{eq:cuculle}\end{equation}

\end{itemize}
\noindent For the sake of completeness, the flowchart of this \emph{GA}-based
design process is reported in Fig. 1(\emph{b}) where the non-standard
operators have been highlighted with a red contour.

\noindent Concerning the computational complexity, it is worth noticing
that (\emph{a}) the cost function in (\ref{eq:AD cost function})
is evaluated in the \emph{AD} and not in the \emph{PD} as (\ref{eq:mask constraint})
with an inexpensive operation without requiring a (fine) sampling
of the power pattern, (\emph{b}) the local search, implemented throughout
the {}``\emph{Post}-\emph{GA} \emph{Cyclic Shift}'', actually requires
(\ref{eq:cuculle}) the computation of the normalized pattern $\widetilde{\mathcal{E}}\left(u;\widehat{\bm{\alpha}}^{\left(\sigma\right)}\right)$
(\ref{eq:normalized pattern}) and the evaluation of its deviation
from the mask $\mathcal{M}\left(u\right)$ (\ref{eq:mask matching})
only for $P$ shifted sequences.

\section{\noindent Numerical Assessment and Validation\label{sec:Numerical-Analysis-and}}

\noindent This section is devoted to assess the proposed \emph{AD}-based
thinning approach and its algorithmic implementations as well as to
derive suitable guidelines for its application. Moreover, the results
from comparisons with conventional \emph{PD}-based methods will be
reported to point out the effectiveness and the efficiency of the
\emph{AD}-driven synthesis paradigm.

\noindent The first numerical analysis is aimed at illustrating the
features of the \emph{AD} formulation in comparison with the traditional
\emph{PD} one. Towards this end, a half-wavelength spaced ($\Delta z=\frac{\lambda}{2}$)
\emph{TIA} composed of $P=16$ ideal radiators and complying with
the following mask

\noindent \begin{equation}
\mathcal{M}\left(u\right)=\left\{ \begin{array}{cc}
0\,\textrm{[dB]}; & -\frac{1.0}{P}<u<\frac{1.0}{P}\\
-15\,\textrm{[dB]}; & otherwise\end{array}\right.\label{eq:simple mask}\end{equation}
has been chosen as benchmark so that it has been possible to \emph{exhaustively}
compute both the \emph{PD} (\ref{eq:mask matching}) and the \emph{AD}
(\ref{eq:AD cost function}) cost functions thanks to the limited
size (i.e., $H=2^{P}$) of the solution space of the whole set of
admissible thinning sequences ($\mathcal{D}\left(\bm{\alpha}\right)$
$=$ \{${\alpha}_{h}$; $h=1,...,H$\}).

\noindent Figure 2 shows the relative frequency of the cost function
values for both \emph{PD} {[}Fig. 2(\emph{a}) - $\xi\left(\bm{\alpha}\right)${]}
and \emph{AD} {[}Fig. 2(\emph{b}) - $\Phi\left(\bm{\alpha}\right)${]}
formulations. From the inset of Fig. 2(\emph{a}), one can notice that
there are several sub-optimal solutions in close proximity to the
optimal one (i.e., $\bm{\alpha}^{opt}=\arg\left\{ \xi\left(\bm{\alpha}\right)=0\right\} $)
in the \emph{PD} case. This outcome points out the complexity of identifying
the actual global optimum $\bm{\alpha}^{opt}$ when adopting the \emph{PD}
formulation, regardless of the solution method. On the contrary, the
occurrence of sub-optimal solutions near to $\bm{\alpha}^{opt}$ (
$\bm{\alpha}^{opt}=\arg\left\{ \Phi\left(\bm{\alpha}\right)=0\right\} $)
{[}see the inset of Fig. 2(\emph{b}){]} significantly reduces in the
\emph{AD} formulation, thus avoiding {}``local minima'' for finding
the global one turns out to be an easier task.

\noindent In the following, the numerical analysis will be devoted
to assess the effectiveness of the iterative approach discussed in
Sect. \ref{sec:Method} to faithfully and efficiently retrieve the
global optimum \emph{TIA} layout. Within this framework, the first
test case is concerned with the synthesis of a $P=24$ \emph{TIA},
the target autocorrelation $\bm{\gamma}^{*}$ ($\bm{\gamma}^{*}$
$\triangleq$ \{$\gamma_{s}^{*}$; $s=0,...,P-1$\}) being the known
three-level function coming from an Almost Difference Set (\emph{ADS})
sequence $\bm{\alpha}^{ADS}$ \cite{Oliveri 2009.a}\cite{Oliveri 2010}.
Such a choice has been done since identifying binary sequences complying
with the unique properties of \emph{ADS}s is usually a complex task
even for small apertures \cite{Oliveri 2011e}. 

\noindent Figure 3 summarizes the results. More specifically, Figure
3(\emph{a}) shows that the autocorrelation function of the best parent
$\widehat{\bm{\alpha}}$ is equal, as expected, to that of the final
design $\bm{\alpha}^{opt}$ and they coincide with the target one,
$\bm{\alpha}^{ADS}$, by proving the effectiveness of the synthesis
strategy to explore the autocorrelation solution space and to identify
the global optimum despite the wide dimensionality ($H=2^{24}$) of
search space $\mathcal{D}\left(\bm{\alpha}\right)$. Moreover, the
plot of the convergence values of the cost function (\ref{eq:AD cost function})
for $V=100$ different runs of the evolutionary strategy, \{$\left.\Phi_{i_{conv}}^{best}\right|_{v}$;
$v=1,...,V$\}, in Fig. 3(\emph{b}) assesses the robustness of the
stochastic solver since only in few $v$-th cases $\left.\Phi_{i_{conv}}^{best}\right|_{v}\ne0$,
the value $\Phi_{i_{conv}}^{best}=0$ being that of a parent sequence
$\widehat{\bm{\alpha}}$ with an autocorrelation function equal to
the target one. Figure 3(\emph{b}) also reports the Hamming distance
between $\bm{\alpha}^{ADS}$ and $\widehat{\bm{\alpha}}$,\begin{equation}
\rho\left(\widehat{\bm{\alpha}},\bm{\alpha}^{ADS}\right)\triangleq\sum_{p=0}^{P-1}\frac{\left|\widehat{\alpha}_{p}-\alpha_{p}^{ADS}\right|}{\alpha_{p}^{ADS}},\label{eq:}\end{equation}
 to highlight that generally the \emph{TIA} sequence found at the
{}``\emph{GA Termination}'' step, $\widehat{\bm{\alpha}}$, is not
equal to $\bm{\alpha}^{ADS}$, but one of its cyclic shifted versions
since $\rho\left(\widehat{\bm{\alpha}},\bm{\alpha}^{ADS}\right)>0$.
Such a result points out a key feature of the \emph{AD} formulation,
that is the reduction of the search space, since the whole set of
shifted versions of a binary thinning sequence shares the same value
of (\ref{eq:AD cost function}). For instance, the parent sequence
$\widehat{\bm{\alpha}}$ in Fig. 3(\emph{c}) is quite different from
$\bm{\alpha}^{ADS}$ in Fig. 3(\emph{d}), but $\Phi\left(\widehat{\bm{\alpha}}\right)=\Phi\left(\bm{\alpha}^{ADS}\right)$.
On the other hand, afterwards it is mandatory to apply the {}``\emph{Post-GA
Cyclic Shift''} to identify, through simple comparisons (\ref{eq:cuculle}),
the best shifted version of the parent sequence $\widehat{\bm{\alpha}}$
(i.e., $\bm{\alpha}^{opt}$) that turns out to be equal to $\bm{\alpha}^{ADS}$.

\noindent In the the next example, the target autocorrelation function
is not \emph{a-priori} known and the \emph{TIA} synthesis is formulated
in terms of pattern-mask constraints, $\mathcal{M}\left(u\right)$.
More specifically, such an experiment is aimed at illustrating the
step-by-step application of the \emph{ME-AD-TIA} approach when dealing
with the $P=24$ layout fulfilling the constant sidelobe pattern (\ref{eq:simple mask})
(Fig. 4). Since the method forces the exact match with the target
mask $\mathcal{M}\left(u\right)$ in the sampling points {[}i.e.,
$\mathcal{E}^{ME}\left(\frac{\lambda}{P\Delta z}k\right)=M_{k}$;
$k=0,...,P-1$ being $\mathcal{E}^{ME}\left(u\right)\triangleq\mathcal{E}\left(u;\bm{\alpha}_{ME}^{opt}\right)${]}
{[}Fig. 4(\emph{a}){]}, the target autocorrelation $\bm{\gamma}_{ME}^{*}$
{[}e.g., here the two-level envelope in Fig. 4(\emph{b}){]} could
be a non-feasible function as confirmed by the behavior of the cost
function (\ref{eq:AD cost function}) vs. the iteration index in Fig.
5(\emph{a}). Indeed, the synthesis process does not reach the condition
$\Phi_{i_{conv}}^{best}=0$. However, the autocorrelation of $\widehat{\bm{\alpha}}^{ME}$,
$\widehat{\bm{\gamma}}^{ME}$ {[}$\widehat{\bm{\gamma}}\triangleq\bm{\gamma}\left(\widehat{\bm{\alpha}}\right)${]},
and the target one, $\bm{\gamma}_{ME}^{*}$, turn out to be very close
{[}see yellow vs. red plots in Fig. 5(\emph{b}){]}. Of course, the
same holds true for all the cyclically-shifted sequences, \{$\widehat{\bm{\alpha}}^{\left(\sigma\right)}$;
$\sigma=0,...,P-1$\}, generated by the parent sequence $\widehat{\bm{\alpha}}^{ME}$
{[}Fig. 6(\emph{a}){]}, which include the optimal one $\bm{\alpha}_{ME}^{opt}$
{[}Fig. 6(\emph{b}){]} having the minimum value of (\ref{eq:mask matching})
{[}Fig. 5(\emph{c}){]} according to (\ref{eq:cuculle}).

\noindent The comparison among the target pattern-mask $\mathcal{M}\left(u\right)$
and the patterns from $\widehat{\bm{\alpha}}^{ME}$ and $\bm{\alpha}_{ME}^{opt}$
in Fig. 6(\emph{c}) highlights several key features of the \emph{ME-AD-TIA}
implementation: (\emph{i}) enforcing a {}``strict'' matching with
$\mathcal{M}\left(u\right)$ may yield to an 'unfeasible' synthesis
problem {[}Fig. 6(\emph{c}){]}; (\emph{ii}) the sidelobes of the synthesized
pattern, $\mathcal{E}^{ME}\left(u\right)$, may exceed the mask, $\mathcal{M}\left(u\right)$
{[}i.e., $\mathcal{E}^{ME}\left(u\right)>\mathcal{M}\left(u\right)${]}
since the matching is enforced at the sampling points \{$u_{k}=\frac{\lambda}{P\Delta z}k$;
$k=0,...,P-1$\} {[}Fig. 6(\emph{c}){]}; (\emph{iii}) despite its
simplicity, the {}``\emph{Post-GA Cyclic Shift}{}`` significantly
improves the \emph{SLL} of the synthesized solution {[}i.e., $SLL^{ME}=-10.92$
{[}dB{]} ($SLL^{ME}\triangleq SLL\left\{ \mathcal{E}^{ME}\left(u\right)\right\} $)
vs. $\widehat{SLL}^{ME}=-6.85$ {[}dB{]} ($\widehat{SLL}^{ME}\triangleq SLL\left\{ \widehat{\mathcal{E}}^{ME}\left(u\right)\right\} $;
$\widehat{\mathcal{E}}^{ME}\left(u\right)\triangleq\mathcal{E}\left(u;\widehat{\bm{\alpha}}^{ME}\right)$)
- Fig. 6(\emph{c}){]}.

\noindent It is worth pointing out that, also in this example of a
wider aperture {[}i.e., $P=24$ vs. $P=16${]}, one can still infer
the same outcomes on the distribution of the local minima drawn from
Fig. 2. Once more, the \emph{AD} formulation turns out to be more
suitable for the synthesis of an optimal \emph{TIA} {[}i.e., Fig.
7(\emph{b}) vs. Fig. 7(\emph{a}){]}.

\noindent The same benchmark scenario has then be treated with the
\emph{FPE-AD-TIA} method. Therefore, the feasible auxiliary excitation
set $\mathbf{w}^{feas}$ has been computed by solving the corresponding
conventional mask-constrained fully-populated array design problem
\cite{Rocca 2016}\cite{Sartori 2013}\cite{DUrso 2007} with the
Matlab subroutine \emph{fmincon} \cite{Rocca 2016}\cite{Sartori 2013}\cite{DUrso 2007}.
The plots of $\mathcal{E}^{feas}\left(u\right)$ and the corresponding
samples, \{$E_{k}^{feas}$; $k=0,...,P-1$\}, show that, unlike the
\emph{ME} approach, the \emph{FPE} implementation does not require
the target pattern values to be equal to the mask ones {[}i.e., $E_{k}^{feas}\leq M_{k}$,
$k=0,...,P-1$ - Fig. 4(\emph{a}) and Fig. 8(\emph{d}){]}. Consequently,
the target autocorrelation $\bm{\gamma}_{FPE}^{*}$ turns out to be
different from that of the \emph{ME} case $\bm{\gamma}_{ME}^{*}$
{[}Fig. 4(\emph{b}){]} even though, also in the \emph{FPE} case, $\widehat{\bm{\gamma}}^{FPE}\ne\bm{\gamma}_{FPE}^{*}$
{[}Fig. 8(\emph{a}){]} such as $\widehat{\bm{\gamma}}^{ME}\ne\bm{\gamma}_{ME}^{*}$
{[}Fig. 5(\emph{b}){]}. The result is that the \emph{FPE} pattern
$\mathcal{E}^{FPE}\left(u\right)$ {[}Fig. 8(\emph{d}){]} better complies
with the pattern mask {[}Fig. 8(\emph{d}) vs. Fig. 6(\emph{c}){]}.
In particular, the \emph{SLL} of the \emph{FPE} layout $\bm{\alpha}_{FPE}^{opt}$
in Fig. 8(\emph{c}) is smaller than the \emph{ME} one {[}i.e., $SLL^{FPE}-SLL^{ME}=-3.54$
{[}dB{]} - Fig. 8(\emph{d}) vs. Fig. 6(\emph{c}){]} and the mask $\mathcal{M}\left(u\right)$
is violated only in a few angular samples {[}Fig. 8(\emph{d}){]}.
As theoretically expected, the \emph{FPE} implementation guarantees,
on the one hand, an efficient solution space exploration thanks to
the \emph{AD} formulation, and on the other hand, an intrinsically
more accurate pattern control than the \emph{ME} technique.

\noindent The dependence of the performance of the \emph{AD}-\emph{TIA}
strategies on the \emph{SLL} of the target pattern mask is analyzed
next by still considering the previous $P=24$ elements benchmark
array. The behavior of $\xi\left(\bm{\alpha}^{opt}\right)$ {[}$\xi_{opt}\triangleq\xi\left(\bm{\alpha}^{opt}\right)${]}
versus the \emph{SLL} value (Fig. 9) in (\ref{eq:simple mask}) confirms
the enhancement of the pattern control enabled by the \emph{FPE} implementation.
As theoretically expected, owing to the limited aperture ($P=24$)
and the binary nature of the \emph{TIA}, the advantage of adopting
the \emph{FPE-AD} method reduces more and more as the \emph{SLL} constraint
gets harder and harder (e.g., $\left.\xi_{opt}^{FPE}\right\rfloor _{SLL=-15\,[\textnormal{dB}]}=2.06\times10^{-4}$
vs. $\left.\xi_{opt}^{ME}\right\rfloor _{SLL=-15\,[\textnormal{dB}]}=7.38\times10^{-3}$;
$\left.\xi_{opt}^{FPE}\right\rfloor _{SLL=-30\,[\textnormal{dB}]}=1.95\times10^{-2}$
vs. $\left.\xi_{opt}^{ME}\right\rfloor _{SLL=-30\,[\textnormal{dB}]}=2.35\times10^{-2}$
- Fig. 9) until unfeasible for the array at hand.

\noindent The fifth test case is concerned with more complex pattern
masks, namely the {}``Tapered Mask'' (\emph{TM}) {[}Fig. 10(\emph{a}){]}
and the {}``Irregular Mask'' (\emph{IM}) {[}Fig. 10(\emph{b}){]}.
The \emph{TIA} layouts synthesized with the \emph{FPE-AD} {[}Figs.
10(\emph{a})-10(\emph{b}){]} turn out to be almost completely compliant
with the corresponding pattern masks $\mathcal{M}\left(u\right)$
{[}Figs. 10(\emph{c})-10(\emph{d}){]}. Moreover, once again the {}``\emph{Post-GA
Cyclic Shift}{}`` significantly improves the mask fitting of the
parent sequence $\widehat{\mathcal{E}}^{FPE}\left(u\right)$ since
$\xi_{opt}^{FPE}<\widehat{\xi}^{FPE}$ {[}$\widehat{\xi}\triangleq\xi\left(\widehat{\bm{\alpha}}\right)${]}:
$\left.\xi_{opt}^{FPE}\right\rfloor _{TM}=1.78\times10^{-4}$ vs.
$\left.\widehat{\xi}^{FPE}\right\rfloor _{TM}=2.57\times10^{-3}$
and $\left.\xi_{opt}^{FPE}\right\rfloor _{IM}=9.86\times10^{-4}$
vs. $\left.\widehat{\xi}^{FPE}\right\rfloor _{IM}=1.52\times10^{-2}$.

\noindent To assess the reliability of the \emph{AD}-based \emph{TIA}
synthesis in real systems where coupling effects arise along with
other non-idealities, the design of an array composed of cavity-backed
slot-fed patch antennas that resonate at $28$ GHz {[}see Fig. 11(\emph{a})
and Tab. I{]} to afford the element pattern in Fig. 11(\emph{b}) has
been carried out by enforcing the same mask profiles of Fig. 10. Figure
12 shows the plots of the 3D {[}Figs. 12(\emph{a})-12(\emph{b}){]}
and the 2D {[}Figs. 12(\emph{c})-12(\emph{d}){]} full-wave \emph{HFSS}
simulated patterns of the \emph{FPE} optimized layouts. The comparisons
between ideal and \emph{HFSS} full-wave simulated patterns in the
elevation cut {[}Figs. 12(\emph{c})-12(\emph{d}){]}, the pattern mask
$\mathcal{M}\left(u\right)$ being also reported, point out the robustness
of the synthesis method, the values of the mask matching error for
the \emph{HFSS} models being very close to the ideal ones {[}i.e.,
$\left.\xi_{opt}^{FPE}\right]_{TM}^{Ideal}=1.78\times10^{-4}$ vs.
$\left.\xi_{opt}^{FPE}\right]_{TM}^{HFSS}=9.42\times10^{-5}$ - Fig.
12(\emph{c}); $\left.\xi_{opt}^{FPE}\right]_{IM}^{Ideal}=9.86\times10^{-4}$
vs. $\left.\xi_{opt}^{FPE}\right]_{IM}^{HFSS}=8.71\times10^{-4}$
- Fig. 12(\emph{d}){]}. 

\noindent The last experiment compares the \emph{FPE-AD} approach
with a traditional \emph{PD}-based one for different apertures. More
specifically, the number of array elements has been varied from $P=16$
up to $P=128$, while enforcing the pattern mask $\mathcal{M}\left(u\right)$
in Fig. 13(\emph{d}) {[}{}``Irregular Mask {[}Type 2{]}'' (\emph{IM2}){]}
and setting the width of the main beam to $\frac{2}{P}$. The plot
of the values of the matching error $\xi$ versus $P$ are shown in
Fig. 13(\emph{a}). As it can be clearly inferred, the \emph{AD}-based
method overcomes the \emph{PD} one regardless of the array aperture
since always $\xi_{opt}^{FPE-AD}<\xi_{opt}^{PD}$ {[}Fig. 13(\emph{a}){]}.
For illustrative purposes, Figure 13(\emph{d}) reports the power patterns
radiated by the $P=48$ array layouts in Figs. 13(\emph{b})-13(\emph{c}).
In this case, $\left.\xi_{opt}^{FPE-AD}\right\rfloor _{P=48}=1.47\times10^{-3}$
, while $\left.\xi_{opt}^{PD}\right\rfloor _{P=48}=8.97\times10^{-3}$
{[}Fig. 13(\emph{a}){]}.

\noindent As for the computational cost, the plots of the synthesis
time, $\Delta t$, versus the array size $P$ are given in Fig. 14(\emph{a}).
It is worthwhile to point out that non-optimized \emph{MATLAB} implementations
have been considered for both \emph{AD}- and \emph{PD}-based formulations,
the SW codes being executed on single-core laptop with $2.4$ {[}GHz{]}
\emph{CPU}. As expected and claimed at the bottom of Sect. \ref{sec:Method},
thanks to the faster cost function computation (\ref{eq:AD cost function})
in the solution-search process, the \emph{AD} synthesis turns out
to be considerably more efficient regardless of the \emph{TIA} aperture
(i.e., $-97.2$ \% $\le$ $\frac{\Delta t^{FPE-AD}-\Delta t^{PD}}{\Delta t^{PD}}$
$\le$ $-93.1$ \% - Fig. 14).

\section{\noindent Conclusions and Final Remarks\label{sec:Conclusions-and-Remarks}}

\noindent An innovative \emph{TIA} synthesis method leveraging on
the theoretical framework developed in \cite{Oliveri 2020} has been
proposed. Unlike traditional \emph{PD}-formulated approaches, the
\emph{TIA} design process has been recast as the problem of matching
a target autocorrelation function derived from the user-defined guidelines
and synthesis-objectives modeled as a pattern mask. Two different
implementations of the \emph{AD}-based formulation of the \emph{TIA}
synthesis problem have been proposed and a binary hybrid optimization
approach has been detailed.

\noindent The numerical validation has shown that

\begin{itemize}
\item \noindent formulating the \emph{TIA} synthesis problem in the \emph{AD},
instead of the \emph{PD}, reduces the complexity of the solution process;
\item unlike \cite{Oliveri 2020}, arbitrary mask constraints can be handled
(Fig. 10);
\item within the \emph{AD} problem formulation, the \emph{FPE} implementation
performs better than the \emph{ME} one in terms of both target-mask
fulfillment (Fig. 9) and control of the peak sidelobes (Fig. 6 vs.
Fig 8);
\item the proposed \emph{TIA} synthesis is robust to mutual coupling effects
(Fig. 12);
\item thanks to the reduction of the local minima, the {}``shrinking''
of the solution space, and the faster cost function computation {[}(\ref{eq:AD cost function})
vs. (\ref{eq:mask matching}){]}, \emph{AD}-based design methods are
much more computationally-efficient than the state-of-the-art \emph{PD}
alternatives (Fig. 14).
\end{itemize}
\noindent Future works, beyond the scope of this paper, will be aimed
at either extending to other classes of unconventional/conventional
arrays beyond \emph{TIA} (e.g., clustered, overlapped, thinned as
well as fully-populated arrangements) the proposed \emph{AD}-based
design method, owing to the generality of the \emph{AD} theory, or
handling various kinds of constraints also different from the pattern
mask ones. Moreover, the design of planar as well as conformal layouts
will be the subject of future works.

\section*{\noindent Acknowledgements}

\noindent This work benefited from the networking activities carried
out in the Project ''ICSC National Centre for HPC, Big Data and Quantum
Computing (CN HPC)'' funded by the European Union - NextGenerationEU
within the PNRR Program (CUP: E63C22000970007), the Project {}``INSIDE-NEXT
- Indoor Smart Illuminator for Device Energization and Next-Generation
Communications'' Funded by the European Union under NextGenerationEU
(CUP: E53D23000990001), the Project {}``AURORA - Smart Materials
for Ubiquitous Energy Harvesting, Storage, and Delivery in Next Generation
Sustainable Environments'' funded by the Italian Ministry for Universities
and Research within the PRIN-PNRR 2022 Program (CUP: E53D23014760001),
the Project {}``Telecommunications of the Future'' (PE00000001 -
program {}``RESTART'', Structural Project 6GWINET) funded by European
Union under the Italian National Recovery and Resilience Plan (NRRP)
of NextGenerationEU (CUP: D43C22003080001), the Project \char`\"{}Telecommunications
of the Future\char`\"{} (PE00000001 - program {}``RESTART'', Focused
Project MOSS) funded by European Union under the Italian National
Recovery and Resilience Plan (NRRP) of NextGenerationEU (CUP: J33C22002880001),
the Project {}``Telecommunications of the Future'' (PE00000001 -
program {}``RESTART'', Structural Project IN) funded by European
Union under the Italian National Recovery and Resilience Plan (NRRP)
of NextGenerationEU (CUP: J33C22002880001), the Project {}``Smart
ElectroMagnetic Environment in TrentiNo - SEME@TN'' funded by the
Autonomous Province of Trento (CUP: C63C22000720003), and the Project
DICAM-EXC (Grant L232/2016) funded by the Italian Ministry of Education,
Universities and Research (MUR) within the {}``Departments of Excellence
2023-2027'' Program (CUP: E63C22003880001). A. Massa wishes to thank
E. Vico for her never-ending inspiration, support, guidance, and help.

\newpage
\section*{FIGURE CAPTIONS}

\begin{itemize}
\item \textbf{Figure 1.} Sketch of (\emph{a}) the \emph{TIA} and (\emph{b})
the flowchart of the \emph{GA}-Based design process.
\item \textbf{Figure 2.} \emph{Proof of Concept} ($P=16$, $\Delta z=0.5\lambda$,
$SLL=-15$ {[}dB{]}) - Relative frequency of the $H$ ($H=2^{P}$)
cost function values: (\emph{a}) \emph{PD} and (\emph{b}) \emph{AD}
formulations.
\item \textbf{Figure 3.} \emph{Illustrative Example} ($P=24$, $\Delta z=0.5\lambda$,
\emph{ADS} Matching) - Plot of (\emph{a}) $\gamma_{s}^{*}$, $\gamma_{s}\left(\widehat{\bm{\alpha}}\right)$,
and $\gamma_{s}\left(\bm{\alpha}^{opt}\right)$ ($s=0,...,P-1$),
(\emph{b}) $\left.\Phi_{i_{conv}}^{best}\right|_{v}$ and $\rho\left(\widehat{\bm{\alpha}}_{v},\bm{\alpha}^{ADS}\right)$
($v=1,...,V$; $V=100$), (\emph{c}) the parent layout , $\widehat{\bm{\alpha}}_{v}$
($v=1$), and (\emph{d}) the final \emph{TIA}, \emph{}$\bm{\alpha}_{v}^{opt}$
($v=1$).
\item \textbf{Figure 4.} \emph{Numerical Validation} ($P=24$, $\Delta z=0.5\lambda$,
Flat Sidelobe Mask: $SLL=-15$ {[}dB{]}) - Plots of (\emph{a}) $\mathcal{M}\left(u\right)$
and $\mathcal{E}^{feas}\left(u\right)$ along with the corresponding
samples at $u=u_{k}$ ($u_{k}\triangleq\frac{\lambda}{P\Delta z}k$),
$k=0,...,P-1$, and (\emph{b}) the normalized target autocorrelation
entries, $\left.\gamma_{s}^{*}\right|^{ME}$ and $\left.\gamma_{s}^{*}\right|^{FPE}$
($s=0,...,P-1$).
\item \textbf{Figure 5.} \emph{Numerical Validation} ($P=24$, $\Delta z=0.5\lambda$,
Flat Sidelobe Mask: $SLL=-15$ {[}dB{]}, \emph{ME-AD} Method) - Plots
of (\emph{a}) the iterative evolution of best cost function value,
$\Phi_{i}^{best}$ ($i=1,...,I$; $I=50$), (\emph{b}) the entries
$\gamma_{s}^{*}$, $\gamma_{s}\left(\widehat{\bm{\alpha}}\right)$,
and $\gamma_{s}\left(\bm{\alpha}^{opt}\right)$ ($s=0,...,P-1$),
and (\emph{c}) the mask matching error of the cyclic sequences \{$\xi\left(\widehat{\bm{\alpha}}^{\left(\sigma\right)}\right)$;
($\sigma=0,...,P-1$)\}.
\item \textbf{Figure 6.} \emph{Numerical Validation} ($P=24$, $\Delta z=0.5\lambda$,
Flat Sidelobe Mask: $SLL=-15$ {[}dB{]}, \emph{ME-AD} Method) - Layouts
of (\emph{a}) the parent sequence, $\widehat{\bm{\alpha}}$, and (\emph{b})
the final \emph{TIA} arrangement, $\bm{\alpha}^{opt}$. Plots of (\emph{c})
$\mathcal{M}\left(u\right)$, $\widehat{\mathcal{E}}\left(u\right)$,
$\mathcal{E}\left(u\right)$ and the corresponding samples at \{$u=u_{k}$;
$k=0,...,P-1$\}.
\item \textbf{Figure 7.} \emph{Numerical Validation} ($P=24$, $\Delta z=0.5\lambda$,
Flat Sidelobe Mask: $SLL=-15$ {[}dB{]}, \emph{ME-AD} Method) - Relative
frequency of the $H$ ($H=2^{P}$) cost function values: (\emph{a})
\emph{PD} and (\emph{b}) \emph{AD} formulations.
\item \textbf{Figure 8.} \emph{Numerical Validation} ($P=24$, $\Delta z=0.5\lambda$,
Flat Sidelobe Mask: $SLL=-15$ {[}dB{]}, \emph{FPE-AD} Method) - Plots
of (\emph{a}) the entries $\gamma_{s}^{*}$, $\gamma_{s}\left(\widehat{\bm{\alpha}}\right)$,
and $\gamma_{s}\left(\bm{\alpha}^{opt}\right)$ ($s=0,...,P-1$).
Layouts of (\emph{b}) the parent sequence, $\widehat{\bm{\alpha}}$,
and (\emph{c}) the final \emph{TIA} arrangement, $\bm{\alpha}^{opt}$.
Plots of (\emph{d}) $\mathcal{M}\left(u\right)$, $\widehat{\mathcal{E}}\left(u\right)$,
$\mathcal{E}\left(u\right)$ and the corresponding samples at \{$u=u_{k}$;
$k=0,...,P-1$\}.
\item \textbf{Figure 9.} \emph{Numerical Validation} ($P=24$, $\Delta z=0.5\lambda$,
Flat Sidelobe Mask, \emph{AD} Formulation) - Behaviour of $\xi\left(\bm{\alpha}^{opt}\right)$
versus \emph{SLL}.
\item \textbf{Figure 10.} \emph{Numerical Validation} ($P=24$, $\Delta z=0.5\lambda$,
\emph{FPE-AD} Method) - Layout of (\emph{a})(\emph{b}) the final \emph{TIA}
arrangement, $\bm{\alpha}^{opt}$, and plots of (\emph{c})(\emph{d})
$\mathcal{M}\left(u\right)$, $\widehat{\mathcal{E}}\left(u\right)$,
and $\mathcal{E}\left(u\right)$ when enforcing (\emph{a})(\emph{c})
the {}``Tapered'' (\emph{TM}) or (\emph{b})(\emph{d}) the {}``Irregular''
(\emph{IM}) sidelobe mask.
\item \textbf{Figure 11.} \emph{Numerical Validation} (\emph{Array Element})
- Picture of (\emph{a}) \emph{}the cavity-backed patch and (\emph{b})
the corresponding embedded element pattern. 
\item \textbf{Figure 12.} \emph{Numerical Validation} ($P=24$, $\Delta z=0.5\lambda$,
\emph{FPE-AD} Method) - Plots of (\emph{a})(\emph{b}) the \emph{HFSS}-simulated
\emph{3D} power pattern and (\emph{c})(\emph{d}) $\mathcal{M}\left(u\right)$
and $\mathcal{E}\left(u\right)$ along the elevation cut when enforcing
(\emph{a})(\emph{c}) the {}``Tapered'' (\emph{TM}) or (\emph{b})(\emph{d})
the {}``Irregular'' (\emph{IM}) sidelobe mask.
\item \textbf{Figure 13.} \emph{Numerical Validation} ($\Delta z=0.5\lambda$,
Irregular Mask {[}Type 2{]}) - Behaviour of (\emph{a}) $\xi\left(\bm{\alpha}^{opt}\right)$
versus the aperture size (i.e., the number of array elements, $P$).
Layout of the \emph{TIA} arrangement synthesized with (\emph{b}) the
\emph{PD}-based Method and (\emph{c}) the \emph{FPE-AD} one, $\bm{\alpha}^{opt}$,
along with (\emph{d}) the plots of $\mathcal{M}\left(u\right)$ and
$\mathcal{E}\left(u\right)$ when $P=48$.
\item \textbf{Figure 14.} \emph{Numerical Validation} (Irregular Mask {[}Type
2{]}) - Synthesis time, $\Delta t$, versus the aperture size (i.e.,
the number of array elements, $P$).
\end{itemize}

\section*{TABLE CAPTIONS}

\begin{itemize}
\item \textbf{Table I.} \emph{Numerical Validation} (\emph{Array Element})
- Geometric and dielectric descriptors of the cavity-backed patch.\newpage

\end{itemize}
\begin{center}~\vfill\end{center}

\begin{center}\begin{tabular}{c}
\includegraphics[%
  width=0.90\columnwidth,
  keepaspectratio]{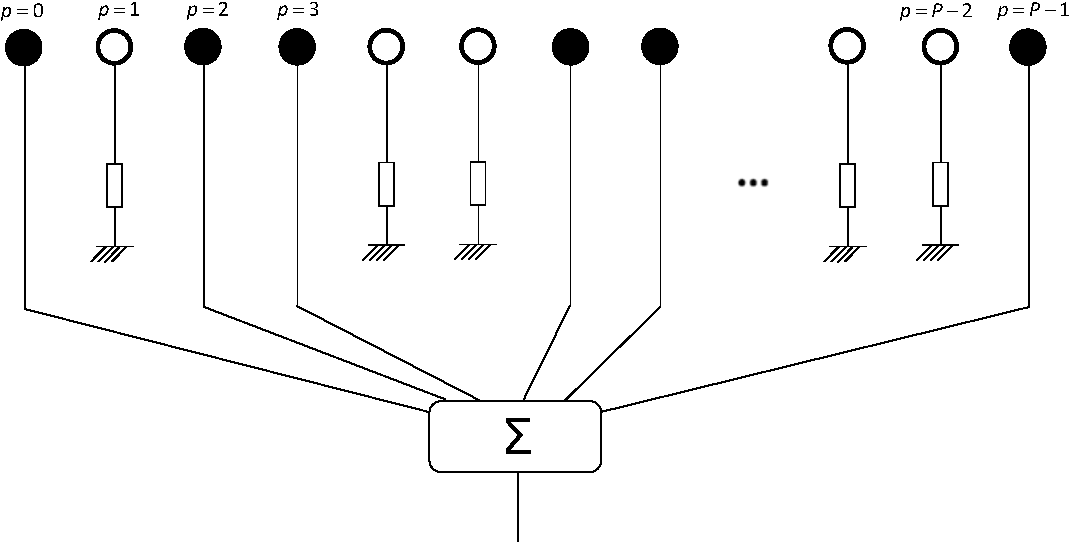}\tabularnewline
(\emph{a})\tabularnewline
\tabularnewline
\includegraphics[%
  width=0.95\columnwidth,
  keepaspectratio]{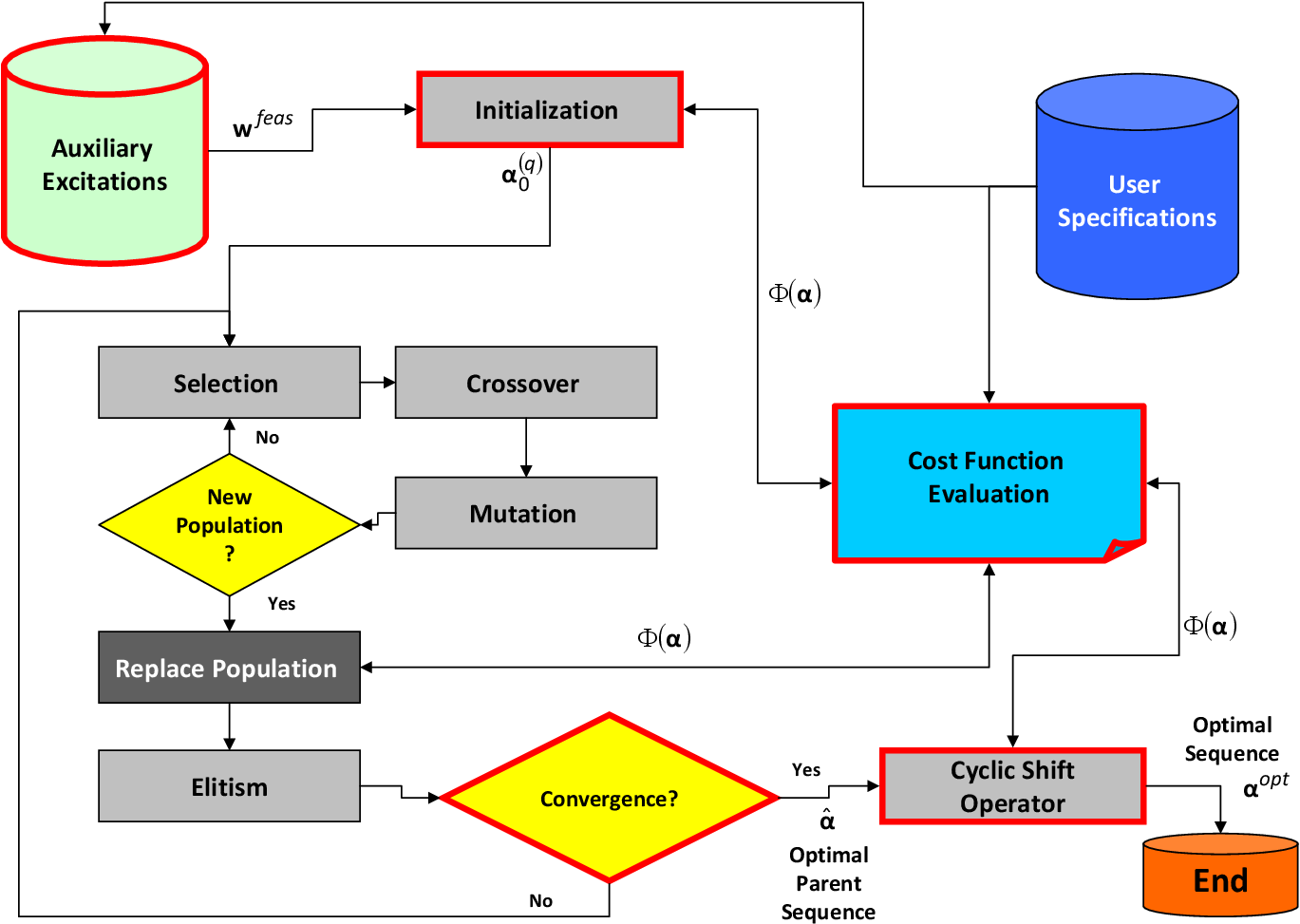}\tabularnewline
(\emph{b})\tabularnewline
\end{tabular}\end{center}

\begin{center}~\vfill\end{center}

\begin{center}\textbf{Fig. 1 - L. Poli et} \textbf{\emph{al.,}} {}``Unconventional
Array Design in the Autocorrelation Domain...''\end{center}

\newpage
\begin{center}~\end{center}

\begin{center}\begin{tabular}{c}
\includegraphics[%
  width=0.90\columnwidth,
  keepaspectratio]{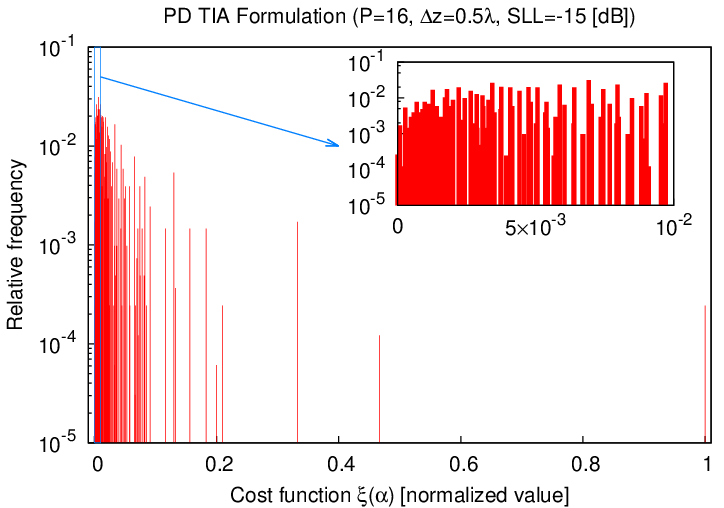}\tabularnewline
(\emph{a})\tabularnewline
\includegraphics[%
  width=0.90\columnwidth,
  keepaspectratio]{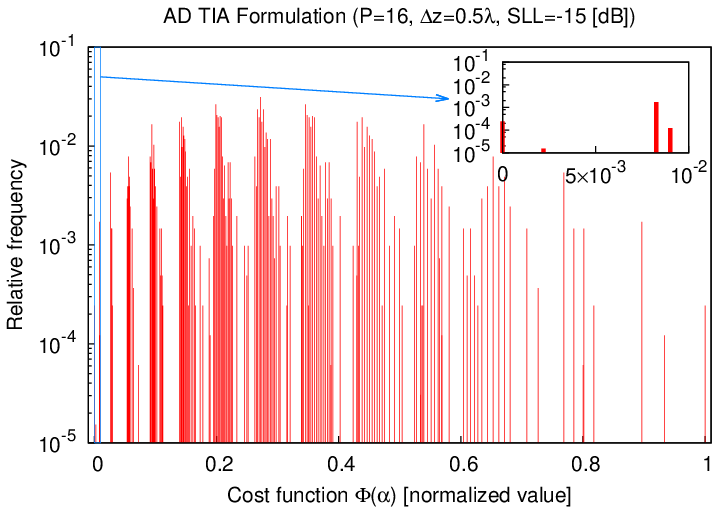}\tabularnewline
(\emph{b})\tabularnewline
\end{tabular}\end{center}

\begin{center}~\vfill\end{center}

\begin{center}\textbf{Fig. 2 - L. Poli et} \textbf{\emph{al.,}} {}``Unconventional
Array Design in the Autocorrelation Domain...''\end{center}
\newpage

\begin{center}\begin{tabular}{c}
\includegraphics[%
  width=0.75\columnwidth,
  keepaspectratio]{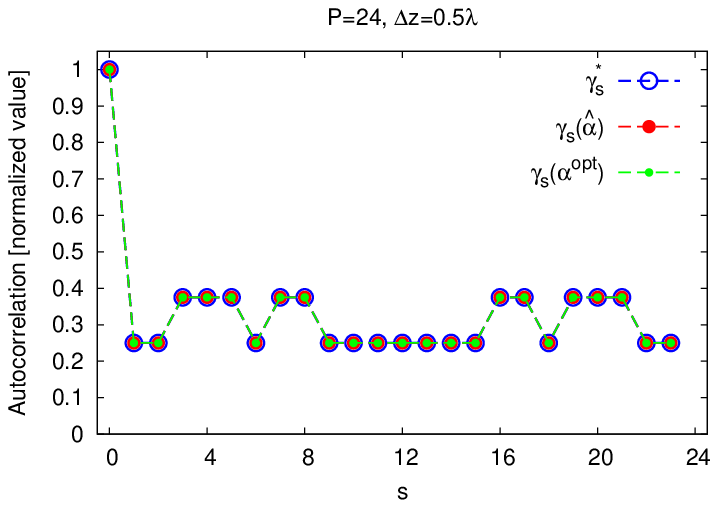}\tabularnewline
(\emph{a})\tabularnewline
\includegraphics[%
  width=0.85\columnwidth,
  keepaspectratio]{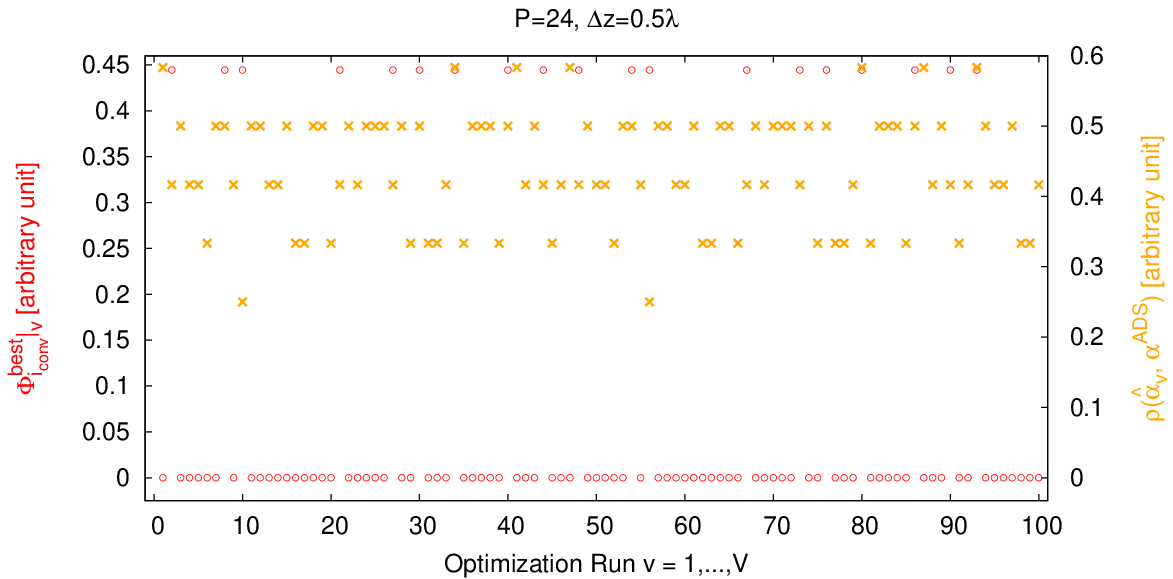}\tabularnewline
(\emph{b})\tabularnewline
\includegraphics[%
  width=0.55\columnwidth,
  keepaspectratio]{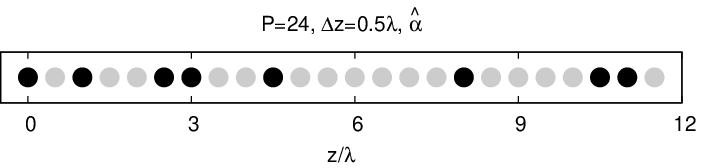}\tabularnewline
(\emph{c})\tabularnewline
\includegraphics[%
  width=0.55\columnwidth,
  keepaspectratio]{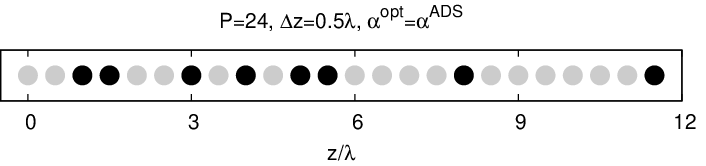}\tabularnewline
(\emph{d})\tabularnewline
\end{tabular}\end{center}

\begin{center}\textbf{Fig. 3 - L. Poli et} \textbf{\emph{al.,}} {}``Unconventional
Array Design in the Autocorrelation Domain...''\end{center}
\newpage

\begin{center}\vfill\end{center}

\begin{center}\begin{tabular}{c}
\includegraphics[%
  width=0.90\columnwidth,
  keepaspectratio]{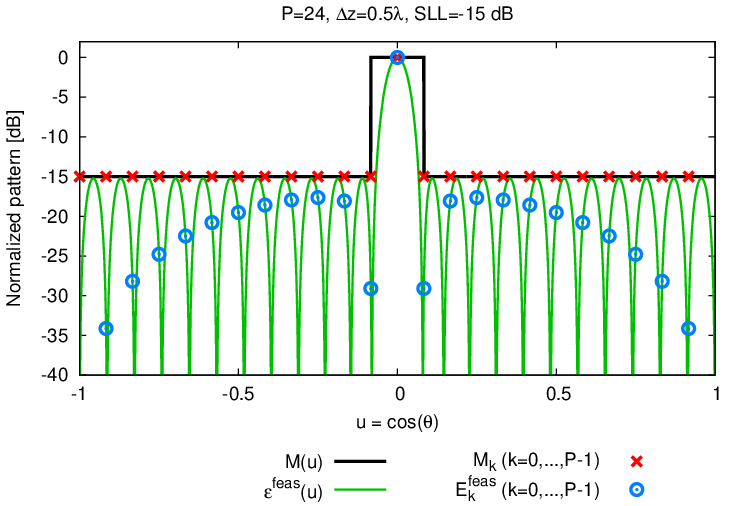}\tabularnewline
(\emph{a})\tabularnewline
\includegraphics[%
  width=0.90\columnwidth,
  keepaspectratio]{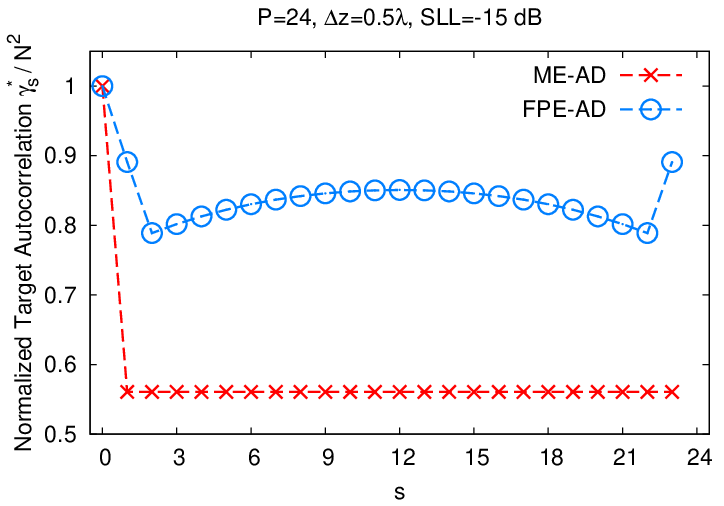}\tabularnewline
(\emph{b})\tabularnewline
\end{tabular}\end{center}

\begin{center}~\vfill\end{center}

\begin{center}\textbf{Fig. 4 - L. Poli et} \textbf{\emph{al.,}} {}``Unconventional
Array Design in the Autocorrelation Domain...''\end{center}
\newpage

\begin{center}\begin{tabular}{c}
\includegraphics[%
  width=0.60\columnwidth,
  keepaspectratio]{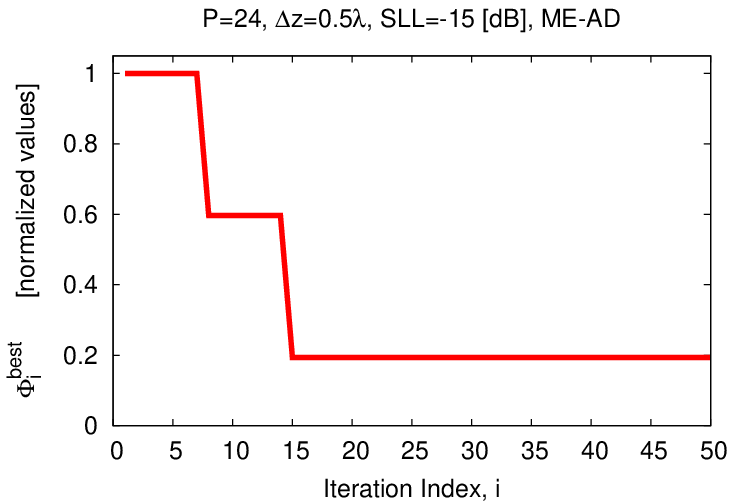}\tabularnewline
(\emph{a})\tabularnewline
\includegraphics[%
  width=0.60\columnwidth,
  keepaspectratio]{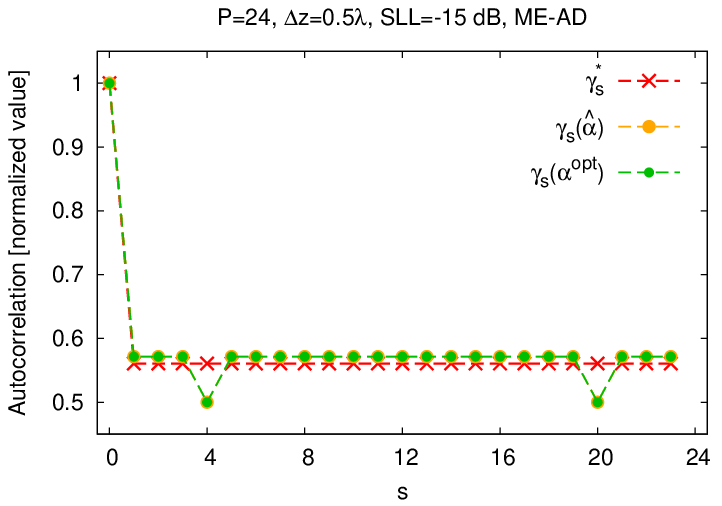}\tabularnewline
(\emph{b})\tabularnewline
\includegraphics[%
  width=0.60\columnwidth,
  keepaspectratio]{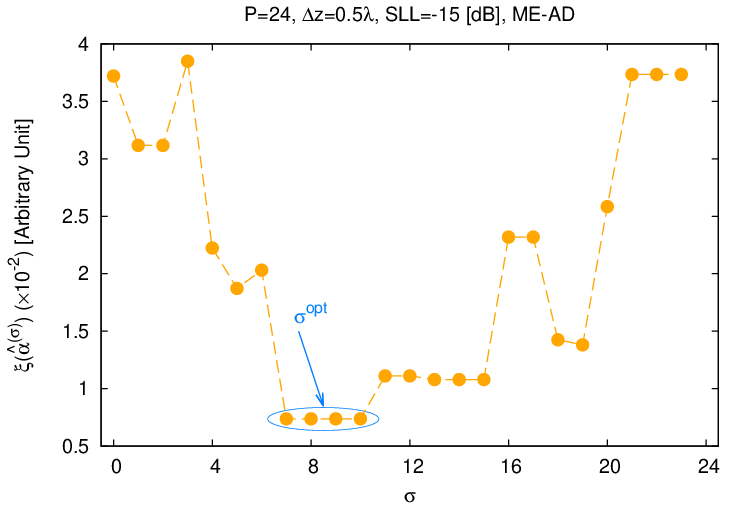}\tabularnewline
(\emph{c})\tabularnewline
\end{tabular}\end{center}

\begin{center}\textbf{Fig. 5 - L. Poli et} \textbf{\emph{al.,}} {}``Unconventional
Array Design in the Autocorrelation Domain...''\end{center}
\newpage

\begin{center}~\vfill\end{center}

\begin{center}\begin{tabular}{c}
\includegraphics[%
  width=0.95\columnwidth,
  keepaspectratio]{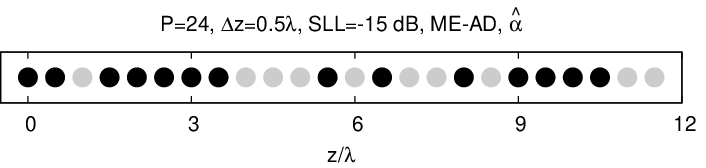}\tabularnewline
(\emph{a})\tabularnewline
\includegraphics[%
  width=0.95\columnwidth,
  keepaspectratio]{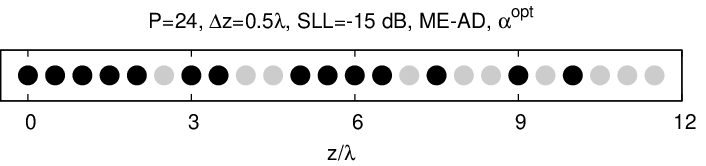}\tabularnewline
(\emph{b})\tabularnewline
\includegraphics[%
  width=0.95\columnwidth,
  keepaspectratio]{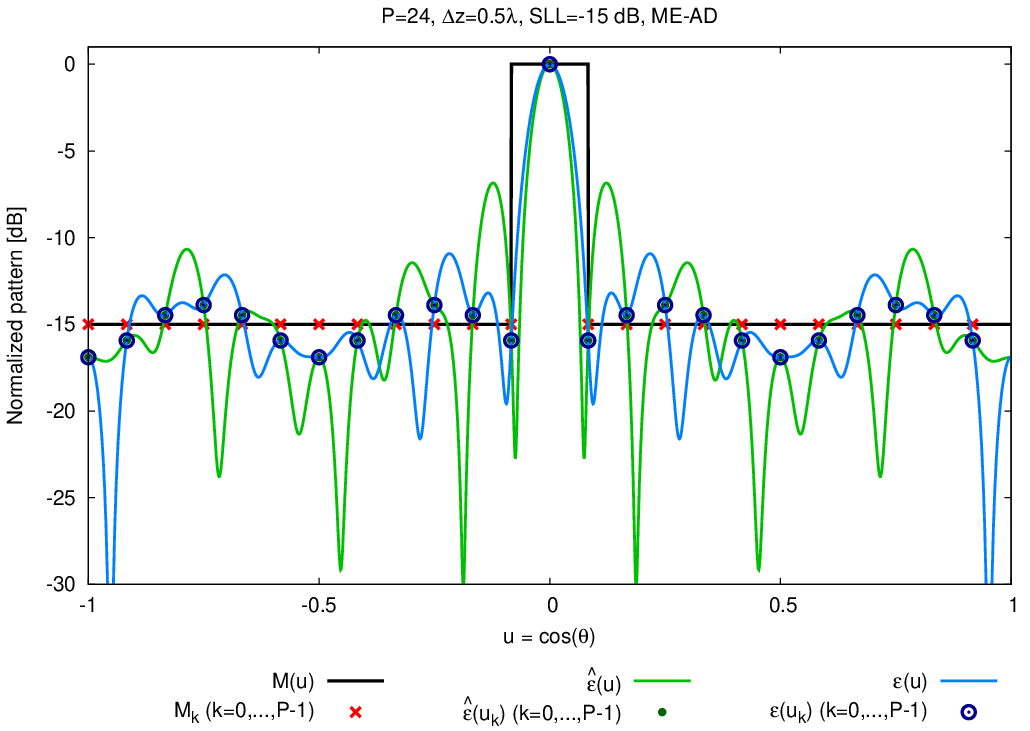}\tabularnewline
(\emph{c})\tabularnewline
\end{tabular}\end{center}

\begin{center}~\vfill\end{center}

\begin{center}\textbf{Fig. 6 - L. Poli et} \textbf{\emph{al.,}} {}``Unconventional
Array Design in the Autocorrelation Domain...''\end{center}
\newpage

\begin{center}~\end{center}

\begin{center}\begin{tabular}{c}
\includegraphics[%
  width=0.90\columnwidth,
  keepaspectratio]{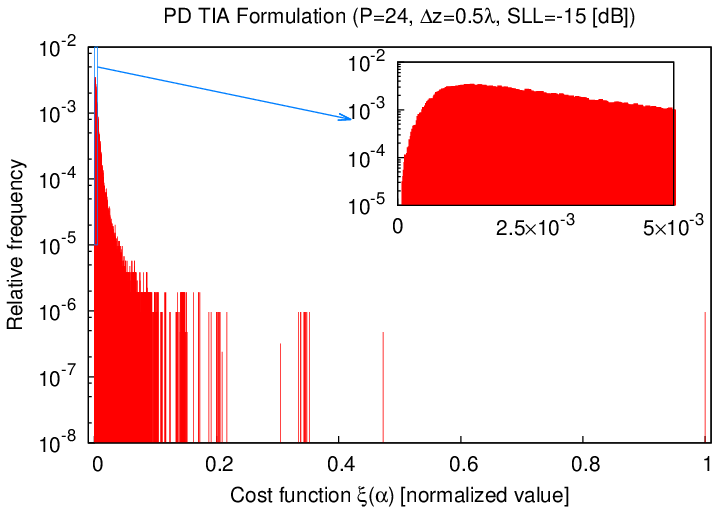}\tabularnewline
(\emph{a})\tabularnewline
\includegraphics[%
  width=0.90\columnwidth,
  keepaspectratio]{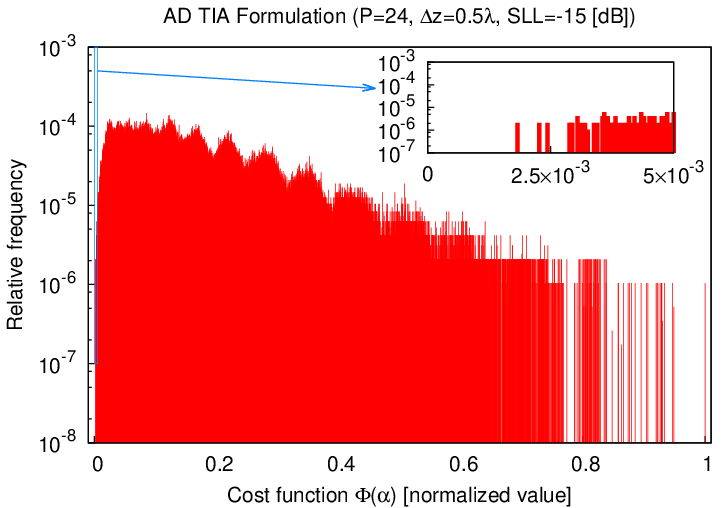}\tabularnewline
(\emph{b})\tabularnewline
\end{tabular}\end{center}

\begin{center}~\vfill\end{center}

\begin{center}\textbf{Fig. 7 - L. Poli et} \textbf{\emph{al.,}} {}``Unconventional
Array Design in the Autocorrelation Domain...''\end{center}
\newpage

\begin{center}\begin{tabular}{c}
\includegraphics[%
  width=0.65\columnwidth,
  keepaspectratio]{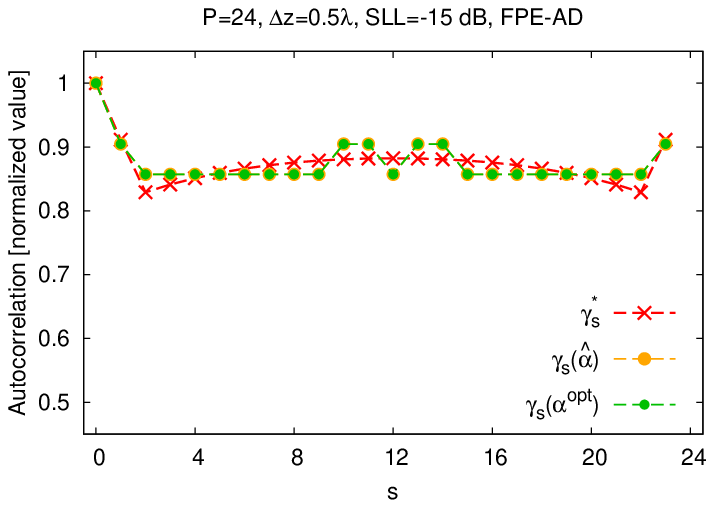}\tabularnewline
(\emph{a})\tabularnewline
\includegraphics[%
  width=0.65\columnwidth,
  keepaspectratio]{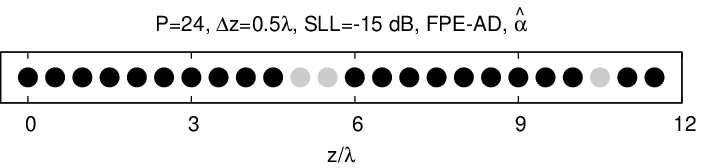}\tabularnewline
(\emph{b})\tabularnewline
\includegraphics[%
  width=0.65\columnwidth,
  keepaspectratio]{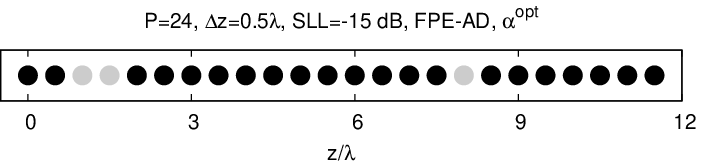}\tabularnewline
(c)\tabularnewline
\includegraphics[%
  width=0.65\columnwidth,
  keepaspectratio]{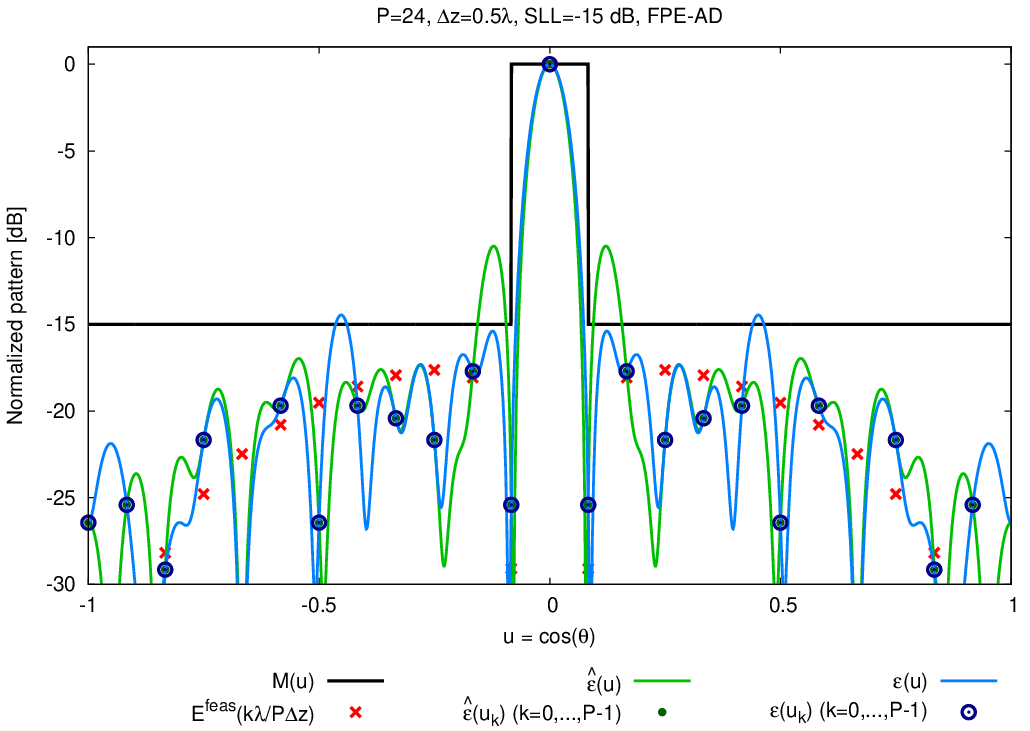}\tabularnewline
(\emph{d})\tabularnewline
\end{tabular}\end{center}

\begin{center}\textbf{Fig. 8 - L. Poli et} \textbf{\emph{al.,}} {}``Unconventional
Array Design in the Autocorrelation Domain...''\end{center}
\newpage

\begin{center}~\vfill\end{center}

\begin{center}\begin{tabular}{c}
\includegraphics[%
  width=0.95\columnwidth,
  keepaspectratio]{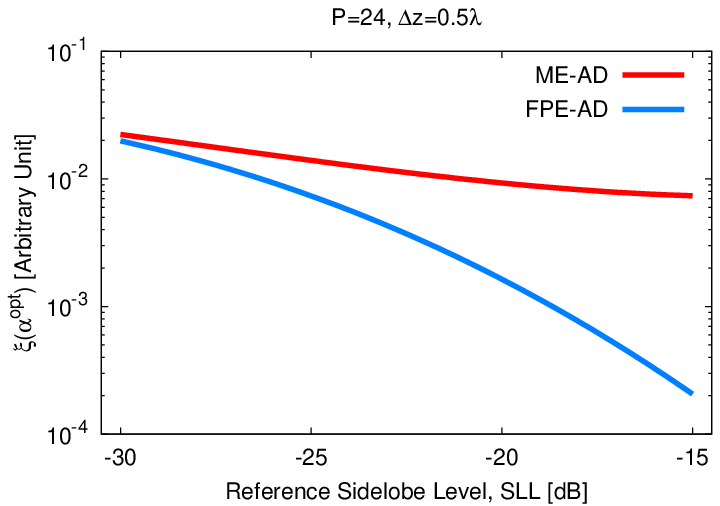}\tabularnewline
\end{tabular}\end{center}

\begin{center}~\vfill\end{center}

\begin{center}\textbf{Fig. 9 - L. Poli et} \textbf{\emph{al.,}} {}``Unconventional
Array Design in the Autocorrelation Domain...''\end{center}
\newpage

\begin{center}~\vfill\end{center}

\begin{center}\begin{tabular}{cc}
\includegraphics[%
  width=0.48\columnwidth,
  keepaspectratio]{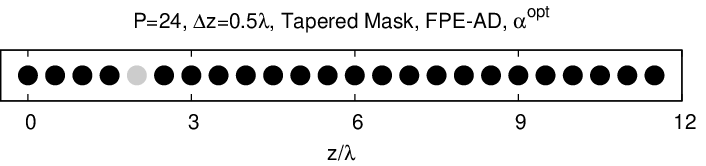}&
\includegraphics[%
  width=0.48\columnwidth,
  keepaspectratio]{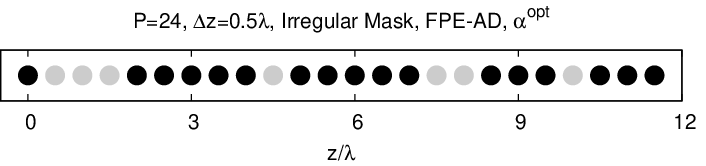}\tabularnewline
(\emph{a})&
(\emph{b})\tabularnewline
\includegraphics[%
  width=0.48\columnwidth,
  keepaspectratio]{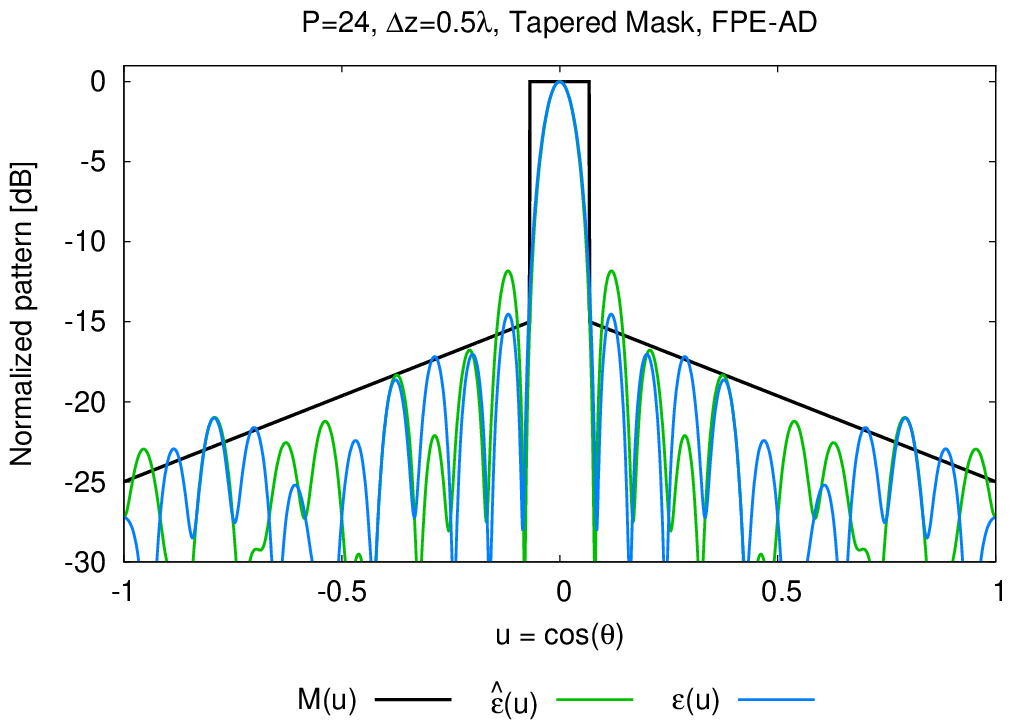}&
\includegraphics[%
  width=0.48\columnwidth,
  keepaspectratio]{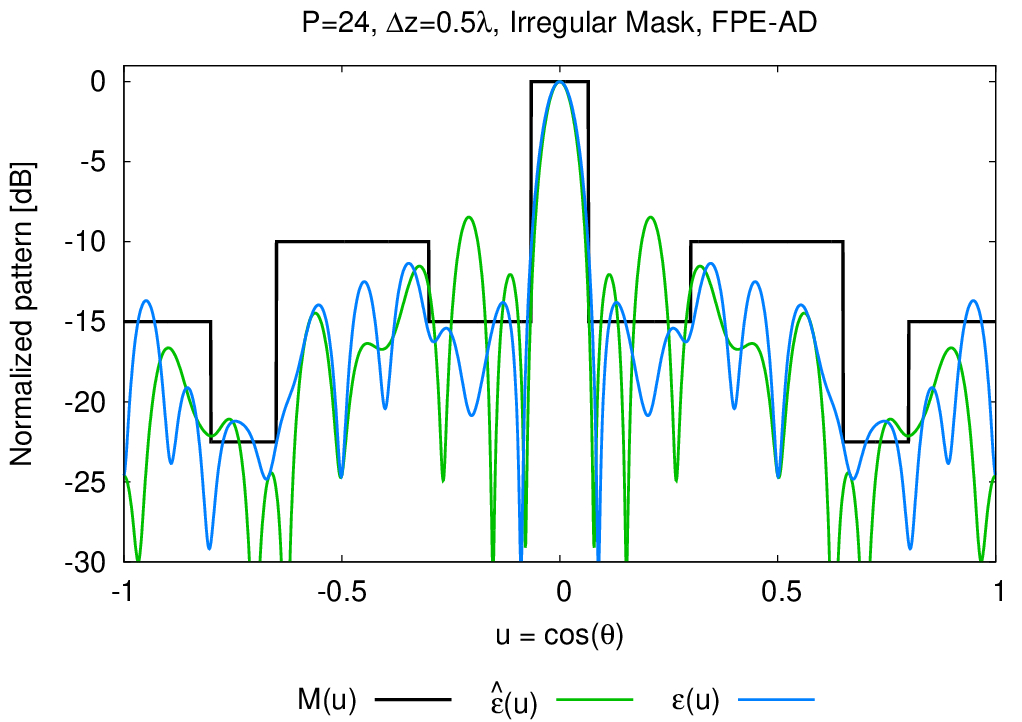}\tabularnewline
(\emph{c})&
(\emph{d})\tabularnewline
\end{tabular}\end{center}

\begin{center}~\vfill\end{center}

\begin{center}\textbf{Fig. 10 - L. Poli et} \textbf{\emph{al.,}} {}``Unconventional
Array Design in the Autocorrelation Domain...''\end{center}
\newpage

\begin{center}~\vfill\end{center}

\begin{center}\begin{tabular}{c}
\includegraphics[%
  width=0.75\columnwidth,
  keepaspectratio]{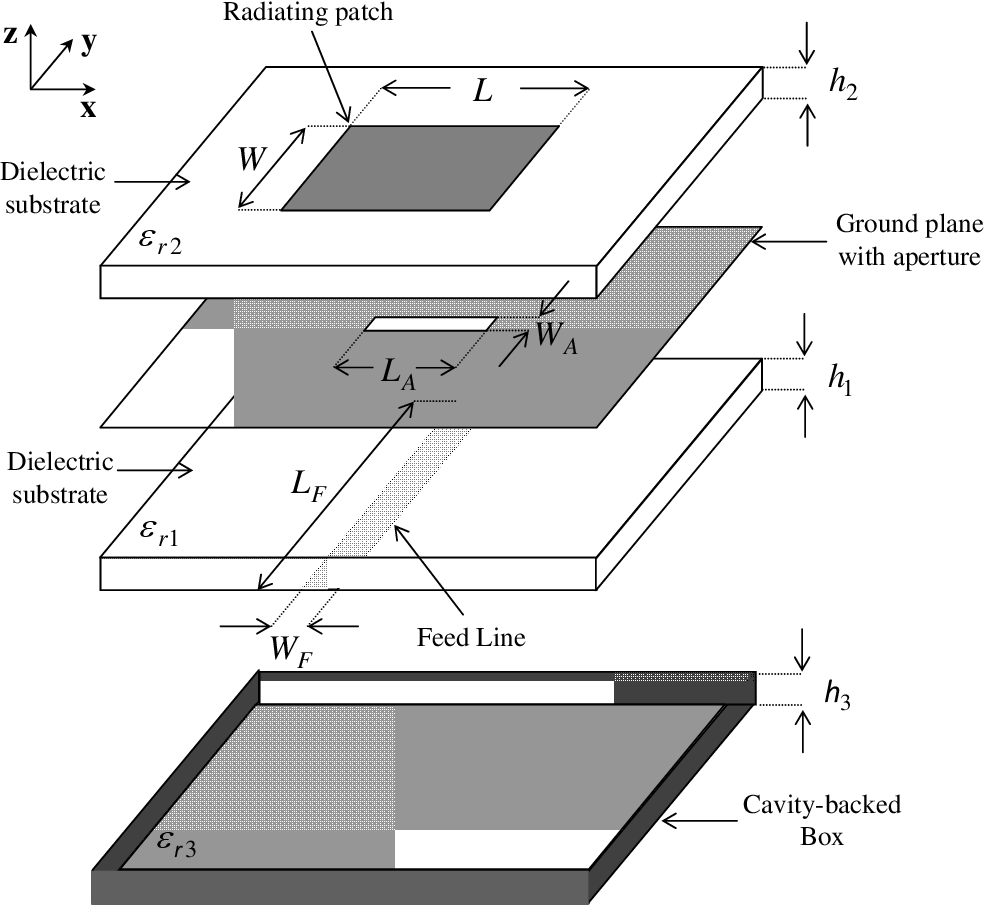}\tabularnewline
(\emph{a})\tabularnewline
\includegraphics[%
  width=0.60\columnwidth,
  keepaspectratio]{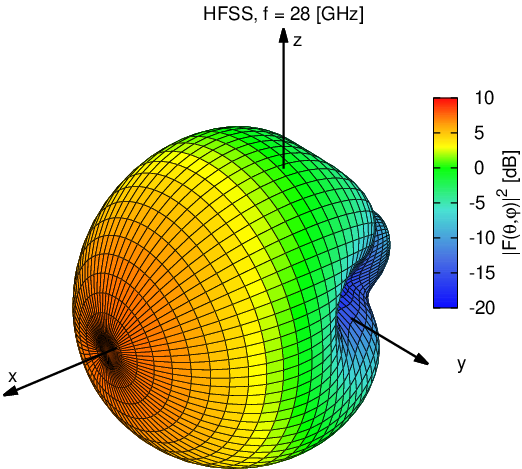}\tabularnewline
(\emph{b})\tabularnewline
\end{tabular}\end{center}

\begin{center}~\vfill\end{center}

\begin{center}\textbf{Fig. 11 -} \textbf{\emph{}}\textbf{L. Poli et}
\textbf{\emph{al.,}} {}``Unconventional Array Design in the Autocorrelation
Domain...''\end{center}
\newpage

\begin{center}~\vfill\end{center}

\begin{center}\begin{tabular}{cc}
\includegraphics[%
  width=0.48\columnwidth,
  keepaspectratio]{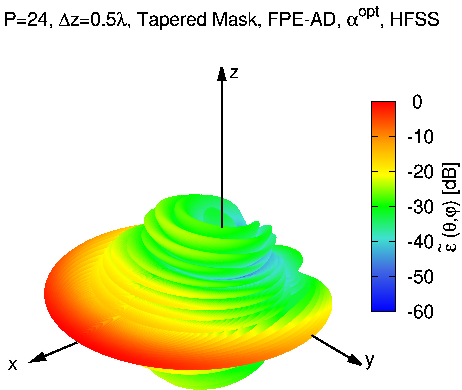}&
\includegraphics[%
  width=0.48\columnwidth,
  keepaspectratio]{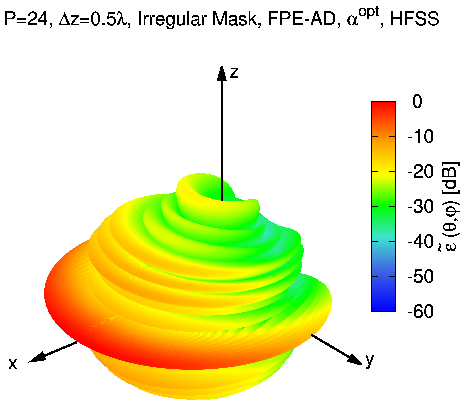}\tabularnewline
(\emph{a})&
(\emph{b})\tabularnewline
\includegraphics[%
  width=0.48\columnwidth,
  keepaspectratio]{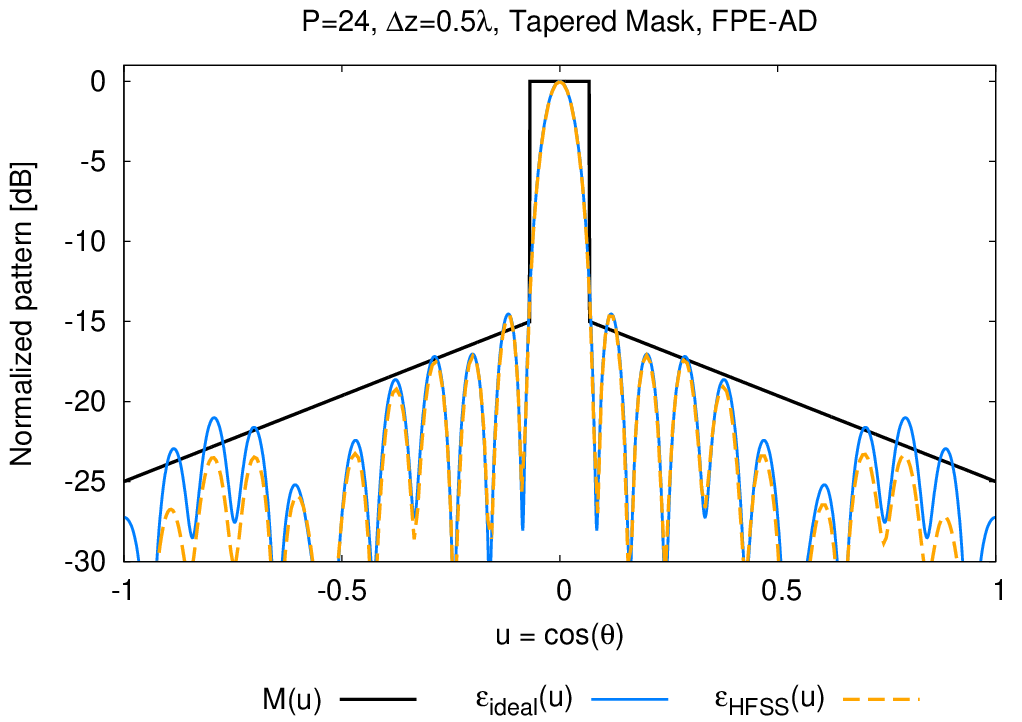}&
\includegraphics[%
  width=0.48\columnwidth,
  keepaspectratio]{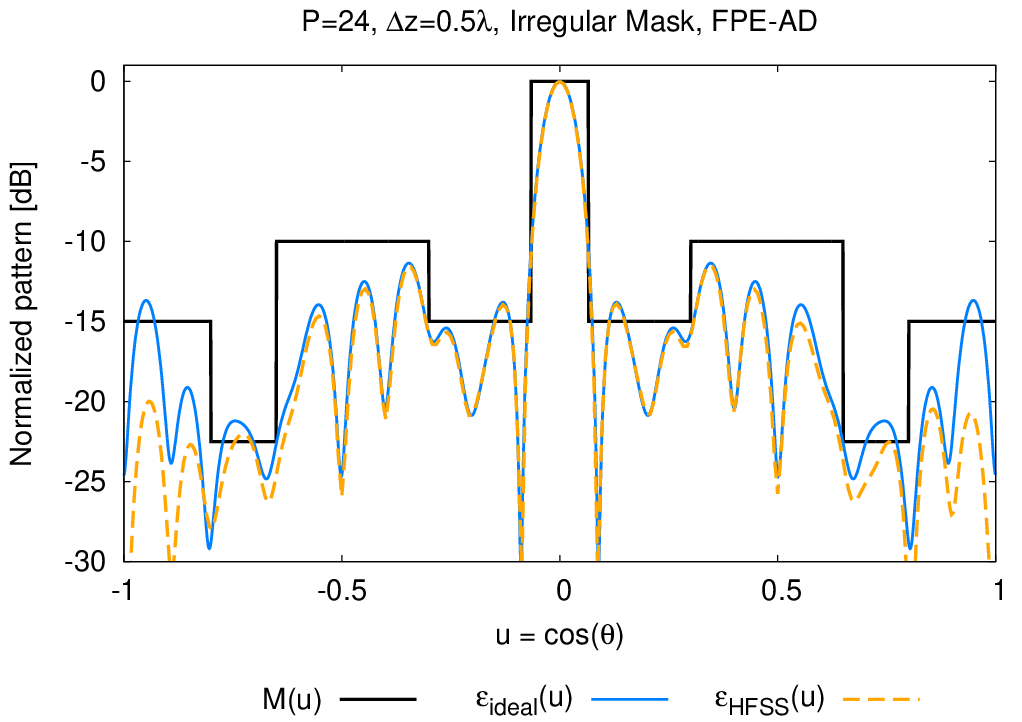}\tabularnewline
(\emph{c})&
(\emph{d})\tabularnewline
\end{tabular}\end{center}

\begin{center}~\vfill\end{center}

\begin{center}\textbf{Fig. 12 -} \textbf{\emph{}}\textbf{L. Poli et}
\textbf{\emph{al.,}} {}``Unconventional Array Design in the Autocorrelation
Domain...''\end{center}
\newpage

\begin{center}\begin{tabular}{c}
\includegraphics[%
  width=0.65\columnwidth,
  keepaspectratio]{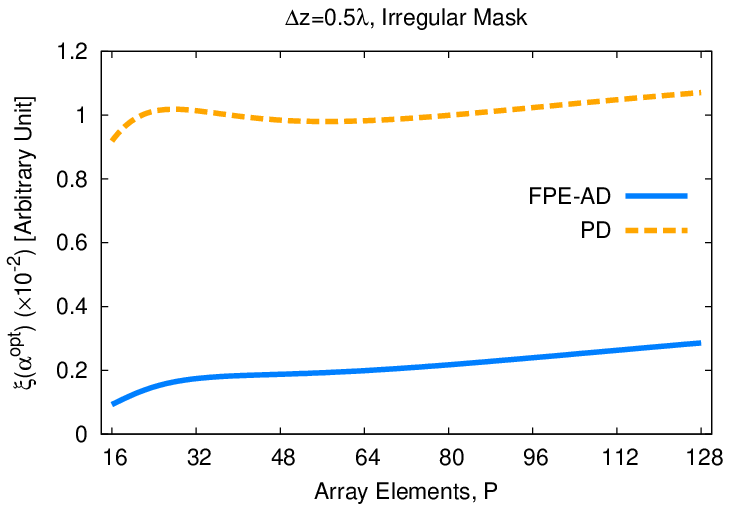}\tabularnewline
(\emph{a})\tabularnewline
\includegraphics[%
  width=0.65\columnwidth,
  keepaspectratio]{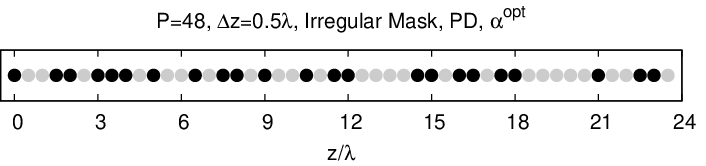}\tabularnewline
(\emph{b})\tabularnewline
\includegraphics[%
  width=0.65\columnwidth,
  keepaspectratio]{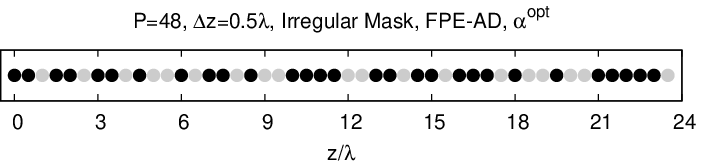}\tabularnewline
(\emph{c})\tabularnewline
\includegraphics[%
  width=0.65\columnwidth,
  keepaspectratio]{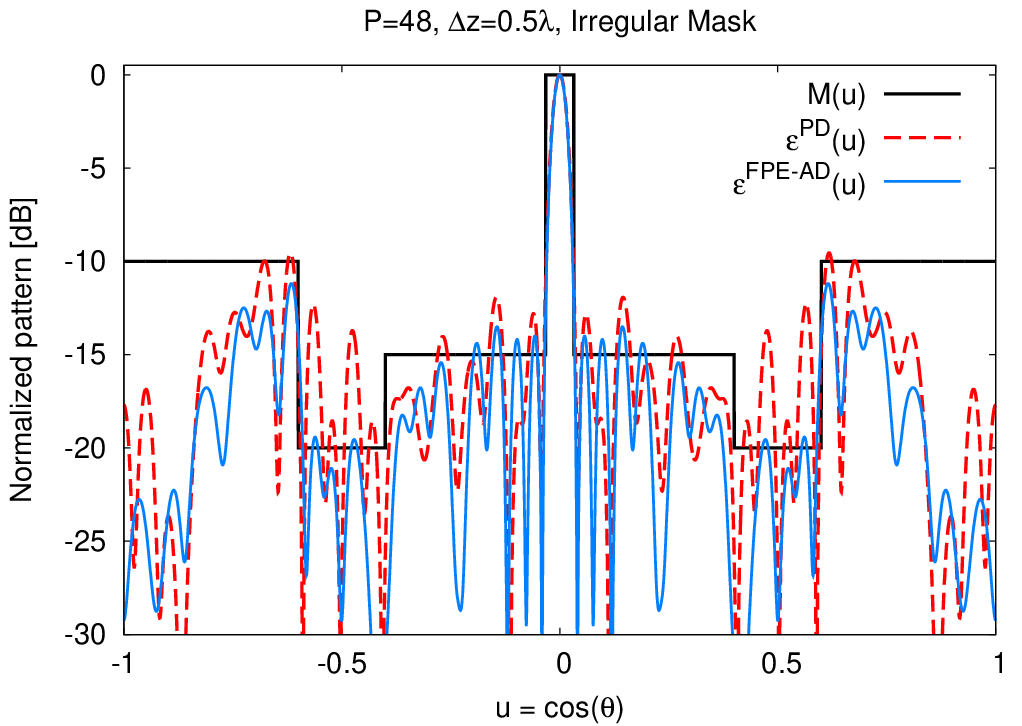}\tabularnewline
(\emph{d})\tabularnewline
\end{tabular}\end{center}

\begin{center}\textbf{Fig. 13 - L. Poli et} \textbf{\emph{al.,}} {}``Unconventional
Array Design in the Autocorrelation Domain...''\end{center}
\newpage

\begin{center}~\vfill\end{center}

\begin{center}\includegraphics[%
  width=0.95\columnwidth,
  keepaspectratio]{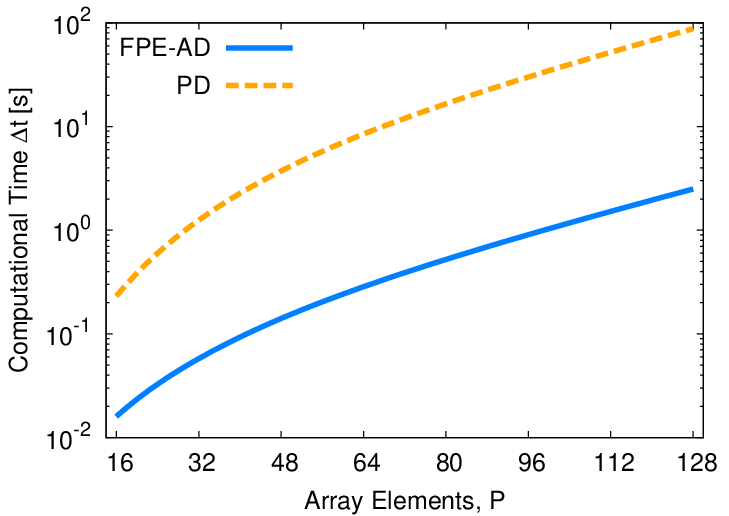}\end{center}

\begin{center}\vfill\end{center}

\begin{center}\textbf{Fig. 14 - L. Poli et} \textbf{\emph{al.,}} {}``Unconventional
Array Design in the Autocorrelation Domain...''\end{center}

\newpage
\begin{center}~\vfill\end{center}

\begin{center}\begin{tabular}{|c|c|c||c|}
\hline 
\multicolumn{3}{|c||}{\textbf{\emph{Geometrical Parameters}} {[}mm{]}}&
\textbf{\emph{Electrical Parameters}}\tabularnewline
\hline
\hline 
$W=4.303$&
$W_{A}=2.151$&
$W_{F}=0.177$&
$\varepsilon_{r1}=2.1$\tabularnewline
\hline
$L=3.260$&
$L_{A}=0.323$&
$L_{F}=2.143$&
$\varepsilon_{r2}=2.1$\tabularnewline
\hline 
$h_{1}=0.627$&
$h_{2}=0.070$&
$h_{3}=1.563$&
$\varepsilon_{r3}=1.0$\tabularnewline
\hline
\end{tabular}\end{center}

\begin{center}~\vfill\end{center}

\begin{center}\textbf{Tab. I - L. Poli et} \textbf{\emph{al.}}\textbf{,}
{}``Unconventional Array Design in the Autocorrelation Domain...''\end{center}\newpage

\end{document}